\documentclass[aps,prd,onecolumn,showpacs,showkeys,amsmath,amssymb]{revtex4}
\usepackage{graphicx}
\usepackage{dcolumn}
\usepackage{bm}
\usepackage[dvips]{color}
\include{shorthand}
\usepackage{amssymb}
\usepackage{amsmath}

\begin{document}

%\begin{document}
\title{Baryon Fields with $U_L(3)\times U_R(3)$ Chiral Symmetry III:
Interactions with Chiral $({\bf 3}, \overline{\bf 3}) \oplus
(\overline{\bf 3}, {\bf 3})$ Spinless Mesons}
\author{Hua-Xing Chen$^{1}$}
\email{hxchen@rcnp.osaka-u.ac.jp}
\author{V. Dmitra\v sinovi\' c$^2$}
\email{dmitrasin@ipb.ac.rs}
\author{Atsushi Hosaka$^{3}$}
\email{hosaka@rcnp.osaka-u.ac.jp}
\affiliation{$^1$Department of Physics and State Key Laboratory of
Nuclear Physics and Technology, Peking University, Beijing 100871,
China \\ $^2$ Institute of Physics, Belgrade University,
Pregrevica 118, Zemun,
P.O.Box 57, 11080 Beograd, Serbia \\
$^3$ Research Center for Nuclear Physics, Osaka University,
Ibaraki 567--0047, Japan}

\begin{abstract}
Three-quark nucleon interpolating fields in QCD have well-defined
$SU_L(3) \times SU_R(3)$ and $U_A(1)$ chiral transformation
properties, {\it viz.} $[({\bf 6},{\bf 3})\oplus({\bf 3},{\bf 6})]$,
$[({\bf 3},\overline{{\bf 3}}) \oplus (\overline{{\bf 3}}, {\bf
3})]$, $[({\bf 8},{\bf 1}) \oplus ({\bf 1}, {\bf 8})]$ and their
``mirror'' images, Ref.~\cite{Chen:2008qv}. It has been shown
(phenomenologically) in Ref.~\cite{Chen:2009sf} that mixing of the
$[({\bf 6},{\bf 3})\oplus({\bf 3},{\bf 6})]$ chiral multiplet with
one ordinary (``naive'') and one ``mirror'' field belonging to the
$[({\bf 3},\overline{{\bf 3}}) \oplus (\overline{{\bf 3}}, {\bf
3})]$, $[({\bf 8},{\bf 1}) \oplus ({\bf 1}, {\bf 8})]$ multiplets
can be used to fit the values of the isovector ($g_A^{(3)}$) and the
flavor-singlet (isoscalar) axial coupling ($g_A^{(0)}$) of the
nucleon and then predict the axial $F$ and $D$ coefficients, or {\it
vice versa}, in reasonable agreement with experiment. In an attempt
to derive such mixing from an effective Lagrangian, we construct all
$SU_{L}(3) \times SU_{R}(3)$ chirally invariant non-derivative
one-meson-baryon interactions and then calculate the mixing angles
in terms of baryons' masses. It turns out that there are (strong)
selection rules: for example, there is only one non-derivative
chirally symmetric interaction between $J=\frac12$ fields belonging
to the $[({\bf 6},{\bf 3})\oplus({\bf 3},{\bf 6})]$ and the $[({\bf
3}, \overline{{\bf 3}}) \oplus (\overline{{\bf 3}}, {\bf 3})]$
chiral multiplets, that is also $U_{A}(1)$ symmetric. We also study
the chiral interactions of the $[({\bf 3}, \overline{{\bf 3}})
\oplus (\overline{{\bf 3}}, {\bf 3})]$ and $[({\bf 8},{\bf 1})
\oplus ({\bf 1}, {\bf 8})]$ nucleon fields. Again, there are
selection rules that allow only one off-diagonal non-derivative
chiral $SU_{L}(3) \times SU_{R}(3)$ interaction of this type, that
also explicitly breaks the $U_A(1)$ symmetry. We use this
interaction to calculate the corresponding mixing angles in terms of
baryon masses and fit two lowest lying observed nucleon (resonance)
masses, thus predicting the third $(J=\frac12, I=\frac32)$ $\Delta$
resonance, as well as one or two flavor-singlet $\Lambda$
hyperon(s), depending on the type of mixing. The effective chiral
Lagrangians derived here may be applied to high density matter
calculations.
\end{abstract}
\maketitle
\pagenumbering{arabic}
%
%===============================================================

\section{Introduction}
\label{Intro}

Axial current ``coupling constants'' of the baryon flavor octet
are well known, see Ref.~\cite{Yamanishi:2007zza}. The zeroth
(time-like) components of these axial currents are generators of
the $SU_L(3) \times SU_R(3)$ chiral symmetry that is one of the
fundamental symmetries of QCD. The general flavor $SU_F(3)$
symmetric form of the nucleon axial current contains two free
parameters, called $F$ and $D$ couplings, that are empirically
determined as $F$=$0.459 \pm 0.008$ and $D$=$0.798 \pm 0.008$, see
Ref.~\cite{Yamanishi:2007zza}. Another, perhaps separate, yet
equally important piece of information is the flavor-singlet axial
coupling $g_A^{(0)}= 0.33 \pm 0.08$ of the
nucleon~\cite{Bass:2007zzb},\cite{Vogelsang:2007zza}.

Recent studies~\cite{Dmitrasinovic:2009vp,Chen:2009sf} point towards
baryon chiral mixing (of $[({\bf 6},{\bf 3})\oplus({\bf 3},{\bf
6})]$ with the $[({\bf 3},\overline{{\bf 3}}) \oplus (\overline{{\bf
3}}, {\bf 3})]$, $[({\bf 8},{\bf 1}) \oplus ({\bf 1}, {\bf 8})]$
chiral multiplets \footnote{These multiplets are not limited to
three-quark interpolators: for a discussion of the validity of our
assumptions, see Sect.~\ref{ssect:summary}.}) as a possible
mechanism underlying the baryons' axial couplings. This finding is
in line with the old current algebra results of Gerstein and
Lee~\cite{Gerstein:1966zz} and of
Harari~\cite{Harari:1966yq,Harari:1966jz}, updated to include
recently measured values of $F$ and $D$ couplings,
Ref.~\cite{Yamanishi:2007zza}, and extended to include the
flavor-singlet coupling $g_A^{(0)}$ of the nucleon, which was not
considered in the mid-1960's at all, presumably due to the lack of
data. Our own starting point was the study of the QCD interpolating
fields' chiral
properties~\cite{Nagata:2007di},\cite{Nagata:2008zzc},\cite{Chen:2008qv}.

The next step is to try and reproduce this phenomenological mixing
starting from a chiral effective model interaction, rather than
{\it per fiat}. As the first step in that direction we must look
for a dynamical source of mixing. One such mechanism is the
simplest chirally symmetric {\it non-derivative}
one-$(\sigma,\pi)$-meson interaction Lagrangian; non-derivative
because that induces baryon masses via the $\sigma$-baryon
coupling.

We construct all $SU_{L}(3) \times SU_{R}(3)$ chirally invariant
non-derivative one-meson-baryon interactions and then use them to
calculate the mixing angles in terms of baryons' masses. It turns
out that there are severe chiral selection rules at work here. For
example, we show that only the mirror field $[(\overline{{\bf 3}},
{\bf 3}) \oplus ({\bf 3},\overline{{\bf 3}})]$ can be coupled to the
$[({\bf 6},{\bf 3})\oplus({\bf 3},{\bf 6})]$ baryon chiral multiplet
by non-derivative terms; whereas the ordinary (``naive") multiplet
$[({\bf 3},\overline{{\bf 3}}) \oplus (\overline{{\bf 3}}, {\bf
3})]$ requires one (or generally an odd number of) derivative(s).
Moreover, this interaction also conserves the $U_A(1)$ symmetry.
This is interesting, as the mixing with a mirror baryon field of
this type seems preferable from the point of view of the two-flavor
phenomenological study, Ref.~\cite{Dmitrasinovic:2009vp}.

We note that all, but one of the $SU_{L}(3) \times SU_{R}(3)$
symmetric interactions, {\it viz.} the $[({\bf 3},\overline{{\bf
3}}) \oplus (\overline{{\bf 3}}, {\bf 3})] -- [({\bf 8},{\bf 1})
\oplus ({\bf 1}, {\bf 8})]$, also conserve the $U_A(1)$ symmetry.
This means that explicit $U_A(1)$ symmetry breaking may occur in
baryons only in so far as the $SU_{L}(3) \times SU_{R}(3)$
symmetry is explicitly broken, with the exception mentioned above.
This is in stark contrast with the $SU_L(2) \times SU_R(2)$ case,
where all of the interaction terms have both the $U_A(1)$
symmetry-conserving and the $U_A(1)$ symmetry-breaking
version~\cite{Dmitrasinovic:2009vp,Dmitrasinovic:2009vy}. In this
sense, the three-flavor chiral symmetry is more restrictive and
consequently more instructive than the two-flavor one.

The conventional models of (linearly realized) chiral $SU_L(3)
\times SU_R(3)$ symmetry,
Refs.~\cite{Hara:1965,Lee:1968,Bardeen:1969ra,Christos2,Zheng:1992mn,Papazoglou:1997uw},
on the other hand appear to fix the $F$ and $D$ parameters at
either ($F$=0,$D$=1), which case goes by the name of $[({\bf
3},\overline{{\bf 3}}) \oplus (\overline{{\bf 3}}, {\bf 3})]$, or
at ($F$=1,$D$=0), which case goes by the name of $[({\bf 8},{\bf
1}) \oplus ({\bf 1}, {\bf 8})]$ chiral representation. Both of
these chiral representations suffer from the shortcoming that
$F$+$D$=1$\neq g_A^{(3)}=$1.267 without derivative couplings. But,
even with derivative interactions, one cannot change the value of
the vanishing coupling, i.e. of $F$=0, in $[({\bf 3},
\overline{{\bf 3}}) \oplus (\overline{{\bf 3}}, {\bf 3})]$, or of
$D$=0, in $[({\bf 8},{\bf 1}) \oplus ({\bf 1}, {\bf 8})]$. Rather,
one can only renormalize the non-vanishing coupling to 1.267. This
is perhaps the most troublesome problem of the linear realization
chiral $SU_L(3) \times SU_R(3)$ symmetric Lagrangians as it has
far-reaching consequences for the kaon and hyperon interactions,
hyper-nuclear physics and nuclear astrophysics of collapsed
stars~\cite{Papazoglou:1998vr,Beckmann:2001bu}.

Another, perhaps equally important and difficult problem is that
of the flavor-singlet axial coupling of the
nucleon~\cite{Bass:2007zzb},\cite{Vogelsang:2007zza}. This is
widely thought of as being disconnected from the $F$,$D$ problem,
but we have already shown, see
Refs.~\cite{Dmitrasinovic:2009vp,Chen:2009sf}, that the chiral
mixing of three-quark interpolating fields casts some new light on
this problem. Namely, the flavor-singlet axial coupling turns out
to be $g_A^{(0)} = (3F-D)$, i.e., a function of the flavor SU(3)
octet $(F,D)$ coefficients and thus proportional to the eighth
flavor component of the $SU(3)$ symmetric axial coupling
$g_A^{(8)}= \frac{1}{\sqrt{3}}(3F-D)$, so long as one mixes only
three-quark interpolating fields. In other words, the ratio of
these two measured quantities is fixed at ${\sqrt{3}}$ in the
three-quark assumption, so one must go beyond this approximation
in order to break the deadlock.

Even though an awareness of this mixing has been around for more
than 40
years~\cite{Hara:1965,Lee:1968,Bardeen:1969ra,Weinberg:1969hw},
the $SU_L(3) \times SU_R(3)$ chiral interactions necessary to
describe such chiral mixing(s) have not been considered in
print~\footnote{D.~Jido and A.~Ohnishi have shown us the results
of some of their unpublished studies~\cite{Jido09,Ohnishi09}. Some
$SU_L(2) \times SU_R(2)$ results can be found in
Refs.~\cite{Dmitrasinovic:2009vp,Dmitrasinovic:2009vy} and some
limited $SU_L(3) \times SU_R(3)$ results can be found in
Refs.~\cite{Christos2},\cite{Zheng:1992mn}.}, let alone derived.
The present paper serves to provide a dynamical model of chiral
mixing that is the ``best" approximation to the phenomenological
solution of both the $(F, D)$ and the flavor-singlet axial
coupling problems, assuming only three-quark baryon interpolating
fields. We found two simple solutions/fits \footnote{which does
not preclude existence of more complicated solutions.}: one that
conserves the $U_A(1)$ symmetry and another one that does not.
This goes to show that the ``QCD $U_A(1)$ anomaly" may, but need
not be the underlying source of the ``nucleon spin problem"
\cite{Bass:2007zzb},\cite{Vogelsang:2007zza}, as was once widely
thought \cite{Zheng:1991pk}. In all likelihood the $U_A(1)$
anomaly provides only a (relatively) small part of the solution,
the largest part coming from the chiral structure (``mixing") of
the nucleon.

One immediate application of our results ought to be in high
density matter calculations, where only one baryon chiral
multiplet ($[({\bf 3},\overline{{\bf 3}}) \oplus (\overline{{\bf
3}}, {\bf 3})]$) and its interaction with mesons have been used
for some time now~\cite{Papazoglou:1998vr,Beckmann:2001bu}.

The present paper consists of five parts: after the present
Introduction, in Sect.~\ref{sect:su3} we define the $SU(3)\times
SU(3)$ chiral transformations of three-quark baryon fields and of
the spinless mesons, with special emphasis on the $SU(3)$ phase
conventions.
%that ensure standard $SU(2)$ isospin conventions for the isospin sub-multiplets,
%and we define the ($SU(3)$ symmetric) vector and axial-vector Noether
%currents of three-quark baryon fields.
In Sect. \ref{sect:interactions} we construct the
%prove the closure of the
$SU_L(3) \times SU_R(3)$ chirally invariant interactions. In
Sect.~\ref{sect:mix} we apply chiral mixing formalism to the
hyperons' axial currents and then use the chiral interactions to
reproduce the mixing angles. In this way we determine the masses
of the admixed states. Finally, in Sect.~\ref{sect:summary} we
discuss the results and offer a summary and an outlook on future
developments.

%===============================================================

\section{Preliminaries: Chiral Transformations of Mesons and
Baryons} \label{sect:su3}

\subsection{Chiral Transformations of $({\bf
3}, \overline{\bf 3}) \oplus (\overline{\bf 3}, {\bf 3})$ Spinless
Mesons} \label{ssect:su3Mesons}

We follow the same definition of chiral transformation in
Ref.~\cite{Chen:2008qv}:
\begin{eqnarray}\nonumber
\bf{U(1)_{V}} &:&  q \to \exp(i {\lambda^0 \over 2} a_{0}) q  = q +
\delta q
\label{eq:u1v} \, , \\
\bf{SU(3)_V} &:& {q} \to \exp (i {\vec \lambda \over 2} \cdot \vec a
){q} =
q + \delta^{\vec{a}} q \label{eq:su3v} \, , \\
\nonumber\bf{U(1)_{A}} &:& q \to \exp(i \gamma_5 {\lambda^0 \over 2}
b_{0}) q = q + \delta_5 q \label{e:u1a} \, , \\ \nonumber
\bf{SU(3)_A} &:& {q} \to \exp (i \gamma_{5} {\vec \lambda \over 2}
\cdot \vec b){q} = q + \delta_5^{\vec{b}} q \label{eq:su3a} \, .
\end{eqnarray}
We define the scalar and pseudoscalar mesons in the $SU(3)$ space:
\begin{eqnarray}
\sigma^a &=& \bar q_A \lambda_{AB}^a q_B \, ,
\\ \pi^a &=& \bar q_A \lambda_{AB}^a i \gamma_5 q_B \, ,
\end{eqnarray}
where the index $a$ goes from 0 to 8, and the zero component of
Gell-Mann matrices is $\lambda^0 = \sqrt{2\over3}{\bf 1}$.

The nucleon fields belong to the chiral representation of $({\bf 3},
\overline{\bf 3}) \oplus (\overline{\bf 3}, {\bf 3})$, and their
combination transforms as:
\begin{eqnarray}
\delta_5^{\vec b} (\sigma^b + i \gamma_5 \pi^b) &=& -  i \gamma_5
b^a d_{abc} (\sigma^c + i \gamma_5 \pi^c) \label{def:D9} \, ,
\\ \nonumber \delta_5^{\vec b} (\sigma^b - i \gamma_5 \pi^b) &=& i \gamma_5
b^a d_{abc} (\sigma^c - i \gamma_5 \pi^c) \, ,
\end{eqnarray}
where $d_{abc}$ and $f_{abc}$ are defined to contain the 0 index:
\begin{equation}
\{ \lambda^a , \lambda^b \} = 2 d^{abc} \lambda^c \, , [ \lambda^a ,
\lambda^b ] = 2 i f^{abc} \lambda^c  \, .
\end{equation}
We note here that in these equations we do not have the
$\delta^{ab}$ factors which are necessary in the usual equation
\begin{equation}
\{ \lambda^a , \lambda^b \} = 2 d^{abc} \lambda^c + {4 \over 3}
\delta^{ab} \, , (a,b = 1 , \cdots 8) \, .
\end{equation}
The nonzero $f$ and $d$ coefficients are:
\begin{eqnarray}
\begin{array}{cc}
abc & f^{abc}
\\ \hline 123 & 1
\\ 147 & 1/2
\\ 156 & -1/2
\\ 246 & 1/2
\\ 257 & 1/2
\\ 345 & 1/2
\\ 367 & -1/2
\\ 458 & \sqrt3/2
\\ 678 & \sqrt3/2
\\ \hline
\end{array}
\hspace{3cm}
\begin{array}{cc|cc|cc}
abc & d^{abc} & abc & d^{abc} & abc & d^{abc}
\\ \hline 000 & \sqrt{2/3} & 118 & {1/\sqrt3} & 355 & 1/2
\\ 011 & \sqrt{2/3} & 146 & 1/2 & 366 & -1/2
\\ 022 & \sqrt{2/3} & 157 & 1/2 & 377 & -1/2
\\ 033 & \sqrt{2/3} & 228 & {1/\sqrt3} & 448 & -1/(2\sqrt3)
\\ 044 & \sqrt{2/3} & 247 & -1/2 & 558 & -1/(2\sqrt3)
\\ 055 & \sqrt{2/3} & 256 & 1/2 & 668 & -1/(2\sqrt3)
\\ 066 & \sqrt{2/3} & 338 & {1/\sqrt3} & 778 & -1/(2\sqrt3)
\\ 077 & \sqrt{2/3} & 344 & 1/2 & 888 & -1/\sqrt3
\\ 088 & \sqrt{2/3} & &
\\ \hline
\end{array}
\end{eqnarray}

To simplify our calculations sometimes we use the ``physical''
basis, whose definitions are:
\begin{eqnarray}
\left (\begin{array}{c} M^1 \\ M^2 \\ M^3 \\ M^4 \\ M^5 \\
M^6 \\ M^7 \\ M^8 \\ M^9
\end{array} \right ) &=& \left (\begin{array}{ccccccccc}
1 & 0 & 0 & 0 & 0 & 0 & 0 & 0 & 0
\\ 0 & {1\over\sqrt2} & -{i\over\sqrt2} & 0 & 0 & 0 & 0 & 0 & 0
\\ 0 & 0 & 0 & 1 & 0 & 0 & 0 & 0 & 0
\\ 0 & {1\over\sqrt2} & {i\over\sqrt2}  & 0 & 0 & 0 & 0 & 0 & 0
\\ 0 & 0 & 0 & 0 & {1\over\sqrt2} & -{i\over\sqrt2} & 0 & 0 & 0
\\ 0 & 0 & 0 & 0 & {1\over\sqrt2} & {i\over\sqrt2} & 0 & 0 & 0
\\ 0 & 0 & 0 & 0 & 0 & 0 & {1\over\sqrt2} & -{i\over\sqrt2} & 0
\\ 0 & 0 & 0 & 0 & 0 & 0 & {1\over\sqrt2} & {i\over\sqrt2} & 0
\\ 0 & 0 & 0 & 0 & 0 & 0 & 0 & 0 & 1
\end{array} \right ) \left (\begin{array}{c} \sigma^0 + i \gamma_5 \pi^0
\\ \sigma^1 + i \gamma_5 \pi^1 \\ \sigma^2 + i \gamma_5 \pi^2 \\ \sigma^3 + i \gamma_5 \pi^3
\\ \sigma^4 + i \gamma_5 \pi^4 \\
\sigma^5 + i \gamma_5 \pi^5 \\ \sigma^6 + i \gamma_5 \pi^6 \\
\sigma^7 + i \gamma_5 \pi^7 \\ \sigma^8 + i \gamma_5 \pi^8
\end{array} \right ) \, .
\end{eqnarray}
In this basis:
\begin{eqnarray}
&& M^1 = \sigma_0 + i \gamma_5 \eta_0 \, , \\ \nonumber && M^2 =
a_0^+ + i \gamma_5 \pi^+ \, , M^3 = a_0^0 + i \gamma_5 \pi^0 \, ,
M^4 = a_0^- + i \gamma_5 \pi^- \, , \\ \nonumber && M^5 = \kappa^+ +
i \gamma_5 K^+ \, , M^6 = \kappa^- + i \gamma_5 K^- \, , M^7 =
\kappa^0 + i \gamma_5 K^0 \, , M^8 = \bar \kappa^0 + i \gamma_5 \bar
K^0 \, , \\ \nonumber && M^9 = f_0 + i \gamma_5 \eta_8  \, .
\end{eqnarray}

We have classified the baryon interpolating fields in our previous
paper~\cite{Chen:2008qv}. We found that the baryon interpolating
fields $N_+^a = N^a_1 + N^a_2$ belong to the chiral representation
$(\mathbf{8}, \mathbf{1}) \oplus (\mathbf{1}, \mathbf{8})$;
$\Lambda$ and $N_-^a = N^a_1 - N^a_2$ belong to the chiral
representation $(\mathbf{3}, \mathbf{\overline{3}}) \oplus
(\mathbf{\overline{3}}, \mathbf{3})$; $N^{a}_\mu$ and
$\Delta^P_{\mu}$ belong to the chiral representation $(\mathbf{6},
\mathbf{3}) \oplus (\mathbf{3}, \mathbf{6})$; and
$\Delta^P_{\mu\nu}$ belong to the chiral representation
$(\mathbf{10}, \mathbf{1}) \oplus (\mathbf{1}, \mathbf{10})$. Here
$N^a_1$ and $N^a_2$ are the two independent kinds of nucleon fields.
$N^a_1$ contains the ``scalar diquark'' and $N^a_2$ contains the
``pseudoscalar diquark''. Moreover, we calculated their chiral
transformations in Ref.~\cite{Chen:2008qv}. In the following
sections, we will use these baryon fields together with one meson
field to construct the chiral invariant Lagrangians.

\subsection{Chiral Transformations of Baryons}
\label{ssect:su3Baryons}

\subsubsection{Chiral Transformations of $[({\bf 6},{\bf 3}) \oplus
({\bf 3},{\bf 6})]$ Baryons}
\label{ssect:(6,3)Baryons}

The baryon field $N_{(18)} = (N_\mu , \Delta_\mu)^T$ belongs to the
chiral representation $[({\bf 6},{\bf 3})\oplus({\bf 3},{\bf 6})]$:
\begin{eqnarray}
&& N^1 = p \, , N^2 = n \, , N^3 = \Sigma^+ \, ,  N^4 = \Sigma^0 \,
, N^5 = \Sigma^- \, ,  N^6 = \Xi^0 \, ,  N^7 = \Xi^- \, ,  N^8 =
\Lambda_8\, ,
\\ \nonumber && N^9 = \Delta^{++} \, , N^{10} = \Delta^{+} \, , N^{11} = \Delta^{0} \, ,
N^{12} = \Delta^{-} \, , \\ \nonumber && N^{13} = \Sigma^{+} \, ,
N^{14} = \Sigma^{0} \, , N^{15} = \Sigma^{-} \, , N^{16} = \Xi^{0}
\, , N^{17} = \Xi^{-} \, ,  N^{18} = \Omega \, ,
\end{eqnarray}
and we can write out their chiral transformation:
\begin{equation}\label{def:F18}
\delta_5^{\vec b} N_{(18)} = i \gamma_5 b^a {\bf F}_{(18)}^a
N_{(18)} = i \gamma_5 b^a \left(
\begin{array}{cc} {\bf D}_{(8)}^a
+ {2\over3}{\bf F}_{(8)}^a & {2\over\sqrt3} {\bf T}_{(8/10)}^a \\
{2\over\sqrt3}{\bf T}^{\dagger a}_{(8/10)} & {1\over3}{\bf
F}_{(10)}^a  \end{array} \right)
\left( \begin{array}{c} N_\mu \\
\Delta_\mu \end{array} \right) \, .
\end{equation}
where the matrices ${\bf D}^a_{(8)}$, ${\bf F}_{(8)}^a$, ${\bf
F}_{(10)}^a$ and ${\bf T}_{(8/10)}^a$ are calculated in our previous
paper~\cite{Chen:2009sf}.

\subsubsection{Chiral Transformations of $[({\bf 3},\overline{{\bf
3}}) \oplus (\overline{{\bf 3}}, {\bf 3})]$ Baryons
}\label{ssect:(3,3)Baryons}

This chiral representation contains the flavor octet and singlet
representations $\mathbf{\bar 3} \otimes \mathbf{3} = \mathbf{8}
\oplus \mathbf{1}$ $\sim N_{(9)} = (\Lambda, N_-)^T$:
\begin{eqnarray}
&& N^1 = \Lambda_0\, , N^2 = p \, , N^3 = n \, , N^4 = \Sigma^+ \, ,
N^5 = \Sigma^0 \, , N^6 = \Sigma^- \, ,  N^7 = \Xi^0 \, , N^8 =
\Xi^- \, ,  N^9 = \Lambda_8\, ,
\end{eqnarray}
and their chiral transformations are
\begin{equation}
\delta_5^{\vec b} N_{(9)} = i \gamma_5 b^a {\bf F}_{(9)}^a N_{(9)}
= i \gamma_5 b^a \left ( \begin{array}{cc} 0 & \sqrt{2\over3}{\bf T}^a_{1/8} \\
\sqrt{2\over3}{\bf T}^{\dagger a}_{1/8} & {\bf D}_{(8)}^a
\end{array} \right ) \left ( \begin{array}{c} \Lambda_1 \\ N_-
\end{array} \right ) \, .
\end{equation}

\subsubsection{Chiral Transformations of $[({\bf 8},{\bf 1})
\oplus ({\bf 1}, {\bf 8})]$ Baryons} \label{ssect:(8,1)Baryons}

This chiral representation $[({\bf 8},{\bf 1}) \oplus ({\bf 1}, {\bf
8})]$ contains the flavor octet representation $\mathbf{8} \otimes
\mathbf{1} = \mathbf{8} $ $\sim N_{(8)} = N_+$. The chiral
transformation is
\begin{eqnarray}
\delta_5^{\vec b} N_{(8)} &=& i \gamma_5 b^a {\rm \bf  F}_{(8)}^a
N_{(8)} \, .
\end{eqnarray}

\subsubsection{Chiral Transformations of $[({\bf 10},{\bf 1})
\oplus ({\bf 1}, {\bf 10})]$ Baryons} \label{ssect:(10,1)Baryons}

This chiral representation $[({\bf 10},{\bf 1}) \oplus ({\bf 1},
{\bf 10})]$ contains the flavor decuplet representation $\mathbf{10}
\otimes \mathbf{1} = \mathbf{10} $ $\sim N_{(10)} =
\Delta_{\mu\nu}$. The chiral transformation is
\begin{eqnarray}
\delta_5^{\vec b} N_{(10)} &=& i \gamma_5 b^a {\rm \bf  F}_{(10)}^a
N_{(10)} \, .
\end{eqnarray}

\section{Chiral Interactions}
\label{sect:interactions}

In this Section we propose a new method for the construction of
$N_f$=3 chiral invariants that differs from the one proposed for
$N_f$=2 in Ref. \cite{Nagata:2008xf} and used in Refs.
\cite{Dmitrasinovic:2009vp,Dmitrasinovic:2009vy}.

\subsection{Diagonal Interactions: Mass Terms}
\label{ssect:diag interaction}

\subsubsection{Chiral $[({\bf 6},{\bf 3})\oplus({\bf 3},{\bf 6})]$
Baryons Diagonal Interactions}
\label{ssect:(6,3)Baryons int}

Our aim is to construct a chiral invariant Lagrangian:
\begin{equation}
\bar N_{(18)}^a M^c N_{(18)}^b {\bf C}^{abc}_{(18)}  \, ,
\label{lag:18}
\end{equation}
where the indices $a$ and $b$ run from 1 to 18, and the index $c$
just runs from 1 to 9. By performing the chiral transformation to
this Lagrangian, we can obtain many equations. For example we have:
\begin{eqnarray}
\delta_5^1 \big (\bar p M^2 n {\bf C}^{122}_{(18)} \big ) &=&
{5\over6} {\bf C}^{122}_{(18)} \bar n  M^2 (i\gamma_5b_1) n + \cdots
\, ,
\\ \nonumber \delta_5^1 \big (\bar \Delta^+ M^2 n {\bf C}^{10,2,2}_{(18)} \big )
&=& -{\sqrt2\over3} {\bf C}^{10,2,2}_{(18)} \bar n M^2
(i\gamma_5b_1) n + \cdots \, ,
\\ \nonumber \delta_5^1 \big (\bar n M^2 \Delta^- {\bf C}^{2,12,2}_{(18)} \big )
&=& \sqrt{2\over3} {\bf C}^{2,12,2}_{(18)} \bar n M^2
(i\gamma_5b_1) n + \cdots \, ,
\\ \nonumber \delta_5^1 \big (\bar n M^1 n {\bf C}^{221}_{(18)} \big ) &=&
{1\over\sqrt{3}} {\bf C}^{221}_{(18)} \bar n M^2 (i\gamma_5b_1) n
+ \cdots \, ,
\\ \nonumber \delta_5^1 \big ( \bar n M^9 n {\bf C}^{229}_{(18)} \big ) &=&
{1\over\sqrt{6}} {\bf C}^{229}_{(18)} \bar n M^2 (i\gamma_5b_1) n
+ \cdots \, .
\end{eqnarray}
These are all the fields that are transformed to $\bar n M^2
(i\gamma_5b_1) n$. If the Lagrangian (\ref{lag:18}) is chiral
invariant, this sum should be zero:
\begin{eqnarray}
{5\over6} {\bf C}^{122}_{(18)} -{\sqrt2\over3} {\bf
C}^{10,2,2}_{(18)} + \sqrt{2\over3} {\bf C}^{2,12,2}_{(18)} +
{1\over\sqrt{3}} {\bf C}^{221}_{(18)} + {1\over\sqrt{6}} {\bf
C}^{229}_{(18)} = 0 \, .
\end{eqnarray}
Solving these equations for ${\bf C}^{abc}_{(18)}$ together with the
hermiticity condition, we find that there is only one solution. The
uniqueness of the solution is guaranteed by the fact that there is
only one way to form the chiral singlet combination out of the
baryon field $[({\bf 6},{\bf 3})\oplus({\bf 3},{\bf 6})]$ and the
meson field $[({\bf 3}, \overline{\bf 3}) \oplus (\overline{\bf 3},
{\bf 3})]$. This solution can be written out much more easily using
${\bf D}_{(18)}^c$ in the following form:
\begin{equation}
g_{(18)} \bar N_{(18)}^a (\sigma^c + i \gamma_5 \pi^c) ({\bf
D}_{(18)}^c)_{ab} N_{(18)}^b \, , \label{def:18 int}
\end{equation}
where $g_{(18)}$ is the coupling constant, and the matrices ${\bf
D}_{(18)}$ are solved to be:
\begin{eqnarray}
{\bf D}^0_{(18)} &=& {1\over\sqrt6} \left ( \begin{array}{cc} {\bf 1}_{8\times8} & 0 \\
0 & -2 \times {\bf 1}_{10\times10}
\end{array} \right ) \, ,
\label{def:D18}
\\ \nonumber {\bf D}^a_{(18)} &=& \left (
\begin{array}{cc} {\rm {\bf D}_{(8)}^{a} +
{2\over3} {\bf F}_{(8)}^{a}} & -{1\over\sqrt3}{\rm {\bf T}_{(8/10)}^a } \\
-{1\over\sqrt3}{\rm {\bf T}^{\dagger a}_{(8/10)} } & -{2\over3}
{\bf F}_{(10)}^{a}
\end{array} \right ) \, .
\end{eqnarray}
Besides the Lagrangian (\ref{lag:18}), its mirror part
\begin{equation}
g_{(18)} \bar N_{(18m)}^a (\sigma^c - i \gamma_5 \pi^c) ({\bf
D}_{(18)}^c)_{ab} N_{(18m)}^b \, , \label{def:18 int2}
\end{equation}
is also chiral invariant. Using these solutions, and performing the
chiral transformation, we can obtain the following relation:
\begin{eqnarray}
{\bf F}^{a\dagger}_{(18)} {\bf D}^b_{(18)} + {\bf D}^b_{(18)} {\bf
F}^a_{(18)} - d_{abc} {\bf D}^c_{(18)} = 0 \, ,
\end{eqnarray}
where ${\bf F}^a_{(18)}$ and ${\bf D}^b_{(18)}$ are defined in the
previous Eqs.~(\ref{def:F18}) and (\ref{def:D18}).

The solution in the physical basis ($\bar N_{(18)}^a M^c N_{(18)}^b
{\bf C}^{abc}_{(18)}$) can be obtained by the following relations:
\begin{eqnarray}\label{eq:twobasis}
&& {\bf C}^{ab1}_{(18)} = ({\bf D}^0_{(18)})_{ab} \, , {\bf
C}^{ab3}_{(18)} = ({\bf D}^3_{(18)})_{ab} \, ,  {\bf
C}^{ab9}_{(18)} = ({\bf D}^8_{(18)})_{ab}  \, ,
\\ \nonumber && {1\over\sqrt2} ({\bf C}^{ab2}_{(18)} + {\bf C}^{ab4}_{(18)}) = ({\bf
D}^1_{(18)})_{ab} \, , {i\over\sqrt2} (-{\bf C}^{ab2}_{(18)} +
{\bf C}^{ab4}_{(18)}) = ({\bf D}^2_{(18)})_{ab}  \, ,
\\ \nonumber && {1\over\sqrt2} ({\bf C}^{ab5}_{(18)} + {\bf C}^{ab6}_{(18)}) = ({\bf
D}^4_{(18)})_{ab} \, , {i\over\sqrt2} (-{\bf C}^{ab5}_{(18)} +
{\bf C}^{ab6}_{(18)}) = ({\bf D}^5_{(18)})_{ab}  \, ,
\\ \nonumber && {1\over\sqrt2} ({\bf C}^{ab7}_{(18)} + {\bf C}^{ab8}_{(18)}) = ({\bf
D}^6_{(18)})_{ab} \, , {i\over\sqrt2} (-{\bf C}^{ab7}_{(18)} +
{\bf C}^{ab8}_{(18)}) = ({\bf D}^7_{(18)})_{ab}   \, .
\end{eqnarray}

\subsubsection{Chiral $[({\bf 3},\overline{{\bf
3}}) \oplus (\overline{{\bf 3}}, {\bf 3})]$ Baryons Diagonal
Interactions} \label{ssect:(3,3)Baryons int}

Following the same procedure of the previous section, we find that
the Lagrangian $\bar N_{(9)}^a M^c N_{(9)}^b {\bf C}^{abc}_{(9)}$
can not be chiral invariant, which means that their is no solution
for ${\bf C}^{abc}_{(9)}$. However, we can still get a chiral
invariant Lagrangian through ``different'' fields. There are two
possible ways:
\begin{enumerate}

\item We use the meson field $\sigma^a - i \gamma_5 \pi^a$:
\begin{eqnarray}
\delta_5^{\vec b} (\sigma^b - i \gamma_5 \pi^b) =  i \gamma_5 b^a
d_{abc} (\sigma^c - i \gamma_5 \pi^c) \, .
\end{eqnarray}

\item We use the mirror field of $N_{(9)}$:
\begin{equation}
\delta_5^{\vec b} N_{(9m)} = - i \gamma_5 b^a {\bf F}_{(9)}^a
N_{(9m)}=
i \gamma_5 b^a \left ( \begin{array}{cc} 0 & -\sqrt{2\over3}{\bf T}^a_{(1/8)} \\
-\sqrt{2\over3}{\bf T}^{\dagger a}_{(1/8)} & -{\bf D}_{(8)}^a
\end{array} \right ) N_{(9m)} \, .
\end{equation}

\end{enumerate}
Then we can construct the chiral invariant Lagrangians:
\begin{equation}\bar N_{(9m)}^a M^c N_{(9m)}^b  {\bf C}^{abc}_{(9)}
\, .
\end{equation}
or its mirror part
\begin{equation}
\bar N_{(9)}^a (M^+)^c  N_{(9)}^b  {\bf C}^{abc}_{(9)} \, ,
\end{equation}
Assuming that they are hermitian, we find that there is only one
solution for ${\bf C}^{abc}_{(9)}$. The solution for the
coefficients ${\bf C}^{abc}_{(9)}$ in these two Lagrangians is the
same, and it can be written out in the following form:
\begin{equation}
g_{(9)}\bar N_{(9m)}^a (\sigma^c + i \gamma_5 \pi^c) ({\bf
D}_{(9)}^c)_{ab} N_{(9m)}^b \label{def:9 int} \, ,
\end{equation}
where the solution is
\begin{eqnarray}
{\bf D}^0_{(9)} &=& {1\over\sqrt6} \left( \begin{array}{cc} -2 &  {\bf 0}_{1\times8} \\
{\bf 0}_{8\times1} & {\bf 1}_{8\times8}
\end{array} \right) \, ,
\\ \nonumber {\bf D}^a_{(9)} &=& \left( \begin{array}{cc} 0  &
{1\over\sqrt6}{\rm {\bf T}^a }_{(1/8)} \\
{1\over\sqrt6}{\rm {\bf T}_{(1/8)}^{\dagger a} } & -\rm {\bf
D}^{a}_{(8)}
\end{array} \right) \, .
\end{eqnarray}
The uniqueness of the solution is guaranteed by the fact that there
is only one way to form the chiral singlet combination out of the
baryon field $[({\bf 3},\overline{{\bf 3}}) \oplus (\overline{{\bf
3}}, {\bf 3})]$ and the meson field $[({\bf 3}, \overline{\bf 3})
\oplus (\overline{\bf 3}, {\bf 3})]$. The coefficients ${\bf
C}^{abc}_{(9)}$ can be similarly obtained like
Eq.~(\ref{eq:twobasis}). From this Lagrangian, we can obtain another
relation:
\begin{eqnarray}
{\bf F}^{a\dagger}_{(9)} {\bf D}^b_{(9)} + {\bf D}^b_{(9)} {\bf
F}^a_{(9)} + d_{abc} {\bf D}^c_{(9)} = 0 \, .
\end{eqnarray}

\subsubsection{Chiral $[({\bf 8},{\bf 1})
\oplus ({\bf 1}, {\bf 8})]$ Baryons Diagonal Interactions}
\label{ssect:(8,1)Baryons int}

Simply adding one $[({\bf 3}, \overline{\bf 3}) \oplus
(\overline{\bf 3}, {\bf 3})]$ meson field to two $[({\bf 8},{\bf 1})
\oplus ({\bf 1}, {\bf 8})]$ baryon fields can not produce a chirally
invariant Lagrangian. By adding two $[({\bf 3}, \overline{\bf 3})
\oplus (\overline{\bf 3}, {\bf 3})]$ meson fields, however, there
are several possible ways to construct chirally invariant
Lagrangians~\cite{Jido09}. First we can write out the group
structures:
\begin{eqnarray}
&&
\big((\mathbf{8},\mathbf{1})\oplus(\mathbf{1},\mathbf{8})\big)^2
\otimes\big((\mathbf{3},\mathbf{\bar 3})\oplus(\mathbf{\bar
3},\mathbf{3})\big)^2
\\ \nonumber &\rightarrow&
\big((\mathbf{1},\mathbf{1})\oplus(\mathbf{1},\mathbf{1})\big)
\otimes
\big((\mathbf{1},\mathbf{1})\oplus(\mathbf{1},\mathbf{1})\big)
\rightarrow
\big((\mathbf{1},\mathbf{1})\oplus(\mathbf{1},\mathbf{1})\big)
---------- (1)
\\ \nonumber &\rightarrow&
\Big(2\times\big((\mathbf{8},\mathbf{1})\oplus(\mathbf{1},\mathbf{8})\big)\Big)
\otimes
\big((\mathbf{8},\mathbf{1})\oplus(\mathbf{1},\mathbf{8})\big)
\rightarrow
2\times\big((\mathbf{1},\mathbf{1})\oplus(\mathbf{1},\mathbf{1})\big)
----- (2)\\ \nonumber &\rightarrow&
\Big(4\times\big((\mathbf{8},\mathbf{8})\oplus(\mathbf{8},\mathbf{8})\big)\Big)
\otimes
\big((\mathbf{8},\mathbf{8})\oplus(\mathbf{8},\mathbf{8})\big)
\rightarrow
4\times\big((\mathbf{1},\mathbf{1})\oplus(\mathbf{1},\mathbf{1})\big)
----- (3)
\end{eqnarray}
Here we just give the Lagrangian for the simplest case (1), which is
$M^{+a} M^a \bar N^b_{(8)} \gamma_5 N^b_{(8m)} + h.c.$. The others
can be obtained by using $M$, $M^+$, $N_{(8)}$ and $N_{(8m)}$ as
well as related coefficients $d_{abc}$ and $f_{abc}$.
%Their respective Lagrangians are
%\begin{enumerate}
%\item[$(1)$] $M_*^aM_*^a \bar N_*^b N_*^b$ \, ,

%\item[$(2)$] $d_{abe} d_{cde} M_*^aM_*^b \bar N_*^c N_*^d~~$ and
%$~~d_{abe} f_{cde} M_*^aM_*^b \bar N_*^c N_*^d$\, ,

%\item[$(3)$] $d_{ace} d_{bde} M_*^aM_*^b \bar N_*^c N_*^d~$, $~~
%d_{ace} f_{bde} M_*^aM_*^b \bar N_*^c N_*^d~$, $~f_{ace} d_{bde}
%M_*^aM_*^b \bar N_*^c N_*^d~$, $~~ f_{ace} f_{bde} M_*^aM_*^b \bar
%N_*^c N_*^d$\, ,
%
%\end{enumerate}
%where $M_*$ denotes either $M$ or $M^+$, and $N_*$ denotes either
%$N_{(8)}$ or $N_{(8m)}$.

\subsubsection{Chiral $[({\bf 10},{\bf 1})
\oplus ({\bf 1}, {\bf 10})]$ Baryons Diagonal Interactions}
\label{ssect:(10,1)Baryons int}

We find that simply adding one $[({\bf 3}, \overline{\bf 3}) \oplus
(\overline{\bf 3}, {\bf 3})]$ meson field to two $[({\bf 10},{\bf
1}) \oplus ({\bf 1}, {\bf 10})]$ baryon fields can not produce a
chirally invariant Lagrangian.

\subsection{Chiral Mixing Interactions}
\label{ssect:off diag interaction}

\subsubsection{Chiral Mixing Interaction
$[({\bf 6},{\bf 3})\oplus({\bf 3},{\bf 6})]$ - $[({\bf 3},
\overline{\bf 3}) \oplus (\overline{\bf 3}, {\bf 3})]$}
\label{ssect:(6,3)(3,3)interaction}

The mixing of $[({\bf 6},{\bf 3})\oplus({\bf 3},{\bf 6})]$ with
$[(\overline{\bf 3}, {\bf 3}) \oplus ({\bf 3}, \overline{\bf 3})]$
(we note that this is a mirror baryon) together a meson field can be
a chiral singlet. So from this section we will study the five
nontrivial off-diagonal Lagrangians.

The simple form made from the ``naive'' baryons $N_{(18)}\sim [({\bf
6},{\bf 3})\oplus({\bf 3},{\bf 6})]$ and $N_{(9)} \sim [({\bf 3},
\overline{\bf 3}) \oplus (\overline{\bf 3}, {\bf 3})]$\, ,
$N^a_{(9)} M^c N_{(18)}^b {\bf C}^{abc}_{(9/18)} + h.c.$ can not be
chiral invariant. We need to use the mirror field $N_{(9m)} \sim
[(\overline{\bf 3}, {\bf 3}) \oplus ({\bf 3}, \overline{\bf
3})]$(mir), and find the following form of field
\begin{eqnarray}
\bar N^a_{(9m)} M^c N^b_{(18)} {\bf C}^{abc}_{(9/18)} + h.c.
\end{eqnarray}
as well as its mirror part can be chiral invariant. Again we turn to
the following form
\begin{eqnarray}
g_{(9/18)} \bar N^a_{(9m)} (\sigma^c + i \gamma_5 \pi^c) ({\bf
T}^c_{(9/18)})_{ab} N^b_{(18)} + h.c. \label{e:9/18 interact}
\end{eqnarray}
We find that the only solution is
\begin{eqnarray}
{\bf T}^0_{(9/18)} &=& {1\over\sqrt6} \left ( \begin{array}{cc}
{\bf 0}_{1\times8} & {\bf 0}_{1\times10} \\
{\bf 1}_{8\times8} & {\bf 0}_{8\times10} \end{array} \right ) \, ,
\\
{\bf T}^a_{(9/18)} &=& \left ( \begin{array}{cc}
-{1\over\sqrt6} {\bf T}^a_{(1/8)} & {\bf 0}_{1\times10} \\
{1\over3}{\bf F}^a_{(8)} & {1\over\sqrt3} {\bf T}^a_{(8/10)}
\end{array} \right )
\label{e:9/18 matrix} \, .
\end{eqnarray}
The coefficients ${\bf C}^{abc}_{(9/18)}$ can be similarly obtained
as in Eq.~(\ref{eq:twobasis}), and we have the following relation:
\begin{eqnarray}
- {\bf F}^{a\dagger}_{(9)} {\bf T}^b_{(9/18)} + {\bf T}^b_{(9/18)}
{\bf F}^a_{(18)} - d_{abc} {\bf T}^c_{(9/18)} = 0 \, .
\end{eqnarray}

\subsubsection{Chiral Mixing Interaction
$[({\bf 6},{\bf 3})\oplus({\bf 3},{\bf 6})]$ -- $[({\bf 8},{\bf
1})\oplus({\bf 1},{\bf 8})]$} \label{ssect:(6,3)(8,1)interaction}

The mixing of a mirror baryon $[({\bf 3},{\bf 6})\oplus({\bf 6},{\bf
3})]$(mir) with $[({\bf 8},{\bf 1})\oplus({\bf 1},{\bf 8})]$
together a meson field can be a chiral singlet, and we find the
following form of field:
\begin{eqnarray}
\bar N^a_{(8)} M^c N^b_{(18m)} {\bf C}^{abc}_{(9/18)} + h.c.
\end{eqnarray}
and its mirror part can be chiral invariant. Again we turn to the
basis
\begin{eqnarray}
g_{(8/18)} \bar N^a_{(8)} (\sigma^c + i \gamma_5 \pi^c) ({\bf
T}^c_{(8/18)})_{ab} N^b_{(18m)} + h.c. \label{e:8/18 interact}
\end{eqnarray}
and the only solution is
\begin{eqnarray}
{\bf T}^0_{(8/18)} &=& {1\over\sqrt6} \left({\bf 1}_{8\times8} ,
{\bf 0}_{8\times10}  \right) \, , \\
{\bf T}^a_{(8/18)} &=& \left( -{1\over2} {\bf D}_{(8)}^a +
{1\over6}{\bf F}_{(8)}^a , -{1\over\sqrt3} {\bf T}^a_{(8/10)}
\right) \label{e:8/18 matrix} \, .
\end{eqnarray}
The coefficients ${\bf C}^{abc}_{(8/18)}$ can be similarly
obtained as in Eq.~(\ref{eq:twobasis}). And we have the following
relation:
\begin{eqnarray}
-{\bf F}^{a\dagger}_{(8)} {\bf T}^b_{(8/18)} + {\bf T}^b_{(8/18)}
{\bf F}^a_{(18)} + d_{abc} {\bf T}^c_{(8/18)} = 0 \, .
\end{eqnarray}

\subsubsection{Chiral Mixing Interaction
$[({\bf 3}, \overline{\bf 3}) \oplus (\overline{\bf 3}, {\bf 3})]$ -
$[({\bf 8},{\bf 1})\oplus({\bf 1},{\bf 8})]$}
\label{ssect:(3,3)(8,1)interaction}

The mixing of $[({\bf 3}, \overline{\bf 3}) \oplus (\overline{\bf
3}, {\bf 3})]$ with $[({\bf 8},{\bf 1})\oplus({\bf 1},{\bf 8})]$
together a meson field can be a chiral singlet, and we find that
there are two possibilities. One is the following form of
Lagrangian:
\begin{eqnarray}
\bar N^a_{(8)} M^c N^b_{(9)} {\bf C}^{abc}_{(8/9)} + h.c.
\end{eqnarray}
and its mirror part can be chiral invariant. Again we turn to the
basis
\begin{eqnarray}
g_{(8/9)} \bar N^a_{(8)} (\sigma^c + i \gamma_5 \pi^c) ({\bf
T}^c_{(8/9)})_{ab} N^b_{(9)} + h.c. \label{e:8/9a interact}
\end{eqnarray}
and the only solution is
\begin{eqnarray}
{\bf T}^0_{(8/9)} &=& {1\over\sqrt6} \left({\bf 0}_{8\times1} , {\bf
1}_{8\times8}
\right) \, , \\
{\bf T}^a_{(8/9)} &=& \left( {1\over\sqrt6} {\bf T}^{\dagger
a}_{(1/8)}, {1\over2} {\bf D}^a_{(8)}+{1\over2} {\bf F}^a_{(8)}
\right) \label{e:8/9a matrix} \, .
\end{eqnarray}
The coefficients ${\bf C}^{abc}_{(8/9)}$ can be similarly obtained
like Eq.~(\ref{eq:twobasis}). and we have the following relation:
\begin{eqnarray}
-{\bf F}^{a\dagger}_{(8)} {\bf T}^b_{(8/9)} - {\bf T}^b_{(8/9)} {\bf
F}^a_{(9)} + d_{abc} {\bf T}^c_{(8/9)} = 0 \, .
\end{eqnarray}

The other possibility is the following form of Lagrangian, and the
mixing of $[({\bf 3}, \overline{\bf 3}) \oplus (\overline{\bf 3},
{\bf 3})]$ with $[({\bf 1},{\bf 8})\oplus({\bf 8},{\bf 1})]$(mir)
\begin{eqnarray}
\bar N^a_{(8m)} M^c N^b_{(9)} {\bf C}^{abc}_{(8/9)} + h.c.
\end{eqnarray}
This and its mirror image part can both be chiral invariant. Again
we turn to the particle basis
\begin{eqnarray}
g_{(B)} \bar N^a_{(8m)} (\sigma^c + i \gamma_5 \pi^c) ({\bf
T}^c_{(B)})_{ab} N^b_{(9)} + h.c. \label{e:8/9b interact}
\end{eqnarray}
The only solution is
\begin{eqnarray}
{\bf T}^0_{B} &=&  {1\over\sqrt6}  \left({\bf 0}_{8\times1} ,
{\bf 1}_{8\times8}  \right) \, , \\
{\bf T}^a_{B} &=& \left( {1\over\sqrt6} {\bf T}^{\dagger a}_{(1/8)},
{1\over2} {\bf D}^a_{(8)}-{1\over2} {\bf F}^a_{(8)} \right)
\label{e:8/9b matrix} \, .
\end{eqnarray}
Since we find that this is the only case which violate the
$U_A(1)$ symmetry, we use the subscript $B$. The coefficients
${\bf C}^{abc}_{(8/9)}$ can be similarly obtained as in
Eq.~(\ref{eq:twobasis}), and we have the following relation:
\begin{eqnarray}
{\bf F}^{a\dagger}_{(8)} {\bf T}^b_{B} - {\bf T}^b_{B} {\bf
F}^a_{(9)} + d_{abc} {\bf T}^c_{B} = 0 \, .
\end{eqnarray}

\subsubsection{Chiral Mixing Interaction
$[({\bf 6},{\bf 3})\oplus({\bf 3},{\bf 6})]$ -
$[({\bf 10},{\bf 1})\oplus({\bf 1},{\bf 10})]$}
\label{ssect:(6,3)(10,1)interaction}

For completeness' sake we also show the $[({\bf 6},{\bf
3})\oplus({\bf 3},{\bf 6})]$ - $[({\bf 10},{\bf 1})\oplus({\bf
1},{\bf 10})]$ chiral mixing interaction. The $[({\bf 10},{\bf
1})\oplus({\bf 1},{\bf 10})]$ decuplet baryon field can only mix
with $[({\bf 3},{\bf 6}) \oplus ({\bf 6},{\bf 3})]$(mir) to compose
a chiral singlet, and we find the following form of Lagrangian:
\begin{eqnarray}
\bar {N}^a_{(10)} M^c N^b_{(18m)} {\bf C}^{abc}_{(10/18)} + h.c.
\end{eqnarray}
and its mirror part can be chiral invariant. Again we turn to the
basis
\begin{eqnarray}
g_{(10/18)} \bar {N}^a_{(10)} (\sigma^c + i \gamma_5 \pi^c) ({\bf
T}^c_{(10/18)})_{ab} N^b_{(18m)} + h.c. \label{e:10/18 interact}
\end{eqnarray}
and the only solution is
\begin{eqnarray}
{\bf T}^0_{(10/18)} &=& {1\over\sqrt6} \left(   {\bf
0}_{10\times8} , {\bf 1}_{10\times10}  \right) \, ,\\
{\bf T}^a_{(10/18)} &=& \left( -{1\over\sqrt3} {\bf T}^{\dagger
a}_{(8/10)}, {1\over3}{\bf F}^a_{(10)} \right) \label{e:10/18
interact} \, .
\end{eqnarray}
The coefficients ${\bf C}^{abc}_{(8/9)}$ can be similarly obtained
like Eq.~(\ref{eq:twobasis}). and we have the following relation:
\begin{eqnarray}
-{\bf F}^{a\dagger}_{(10)} {\bf T}^b_{(10/18)} + {\bf T}^b_{(10/18)}
{\bf F}^a_{(18)} + d_{abc} {\bf T}^c_{(10/18)} = 0 \, .
\end{eqnarray}

\subsection{Brief Summary of Interactions}
\label{ssect:summary}

Altogether we have the following form of chiral invariant
Lagrangian:
\begin{eqnarray}
\label{eq:lag1} \mathcal{L} \nonumber &=& \left( \begin{array}{cccc}
\overline{N}_{(8m)} & \overline{N}_{(9m)} & \overline{N}_{(18)} &
\overline{N}_{(10m)}
\end{array} \right)
\Bigg( (\sigma^a + i \gamma_5 \pi^a) \left (
\begin{array}{cccc}
{\bf 0}_{8\times8} & {\bf 0}_{8\times9} & {\bf 0}_{8\times18} &
{\bf 0}_{8\times10} \\
{\bf 0}_{9\times8} & g_{(9)}{\bf D}^a_{(9)} & g_{(9/18)}{\bf
T}^a_{(9/18)}& {\bf 0}_{9\times10} \\
{\bf 0}_{18\times8} & g^*_{(9/18)}{\bf T}^{\dagger a}_{(9/18)} &
g_{(18/18)}{\bf D}^a_{(18)}& {\bf 0}_{18\times10} \\
{\bf 0}_{10\times8} & {\bf 0}_{10\times9} & {\bf 0}_{10\times18} &
{\bf 0}_{10\times10} \end{array} \right) \\
&& + (\sigma^a - i \gamma_5 \pi^a) \left (
\begin{array}{cccc}
{\bf 0}_{8\times8} & g_{(8/9)}{\bf T}^a_{(8/9)} & g_{(8/18)}{\bf
T}^a_{(8/18)} & {\bf 0}_{8\times10} \\
g_{(8/9)}^*{\bf T}^{\dagger a}_{(8/9)} & {\bf 0}_{9\times9} &
{\bf 0}_{9\times18} & {\bf 0}_{9\times10} \\
g_{(8/18)}^*{\bf T}^{\dagger a}_{(8/18)} & {\bf 0}_{18\times9} &
{\bf 0}_{18\times18} & g_{(10/18)}^*{\bf T}^{\dagger a}_{(10/18)} \\
{\bf 0}_{10\times8} & {\bf 0}_{10\times9} & g_{(10/18)}{\bf
T}^a_{(10/18)} & {\bf 0}_{10\times10}
\end{array} \right)
\Bigg) \left ( \begin{array}{c} {N}_{(8m)} \\
{N}_{(9m)} \\ {N}_{(18)} \\ {N}_{(10m)}
\end{array} \right) \, ,
\end{eqnarray}
its mirror part is also chiral invariant:
\begin{eqnarray}\label{eq:lag2}
\mathcal{L}_{(m)} \nonumber &=& \left ( \begin{array}{cccc}
\overline{N}_{(8)} & \overline{N}_{(9)} & \overline{N}_{(18m)} &
\overline{N}_{(10)}
\end{array} \right) \Bigg( (\sigma^a - i \gamma_5 \pi^a) \left(
\begin{array}{cccc}  {\bf 0}_{8\times8} & {\bf 0}_{8\times9} &
{\bf 0}_{8\times18} & {\bf 0}_{8\times10} \\
{\bf 0}_{9\times8} & g^\prime_{(9)}{\bf D}^a_{(9)} &
g^\prime_{(9/18)}{\bf T}^a_{(9/18)}& {\bf 0}_{9\times10} \\
{\bf 0}_{18\times8} & g^{\prime*}_{(9/18)}{\bf T}^{\dagger
a}_{(9/18)} & g^\prime_{(18/18)}{\bf D}^a_{(18)}& {\bf
0}_{18\times10} \\
{\bf 0}_{10\times8} & {\bf 0}_{10\times9} & {\bf 0}_{10\times18} &
{\bf 0}_{10\times10} \end{array} \right) \\
\nonumber && + (\sigma^a + i \gamma_5 \pi^a) \left(
\begin{array}{cccc}  {\bf 0}_{8\times8} & g^\prime_{(8/9)}{\bf T}^a_{(8/9)} &
g^\prime_{(8/18)}{\bf T}^a_{(8/18)} & {\bf 0}_{8\times10} \\
g_{(8/9)}^{\prime*}{\bf T}^{\dagger a}_{(8/9)} & {\bf
0}_{9\times9} & {\bf 0}_{9\times18} & {\bf 0}_{9\times10} \\
g_{(8/18)}^{\prime*}{\bf T}^{\dagger a}_{(8/18)} & {\bf
0}_{18\times9} & {\bf 0}_{18\times18} & g_{(10/18)}^{\prime*}{\bf
T}^{\dagger a}_{(10/18)} \\
{\bf 0}_{10\times8} & {\bf 0}_{10\times9} & g^\prime_{(10/18)}{\bf
T}^a_{(10/18)} & {\bf 0}_{10\times10} \end{array} \right )
\Bigg ) \left ( \begin{array}{c} {N}_{(8)} \\
{N}_{(9)} \\
{N}_{(18m)} \\ {N}_{(10)} \end{array} \right ) \, .
\end{eqnarray}
Besides these, there is another single piece of Lagrangian which
is also chiral invariant:
\begin{eqnarray}\label{eq:lag1}
\mathcal{L}_{(B)} \nonumber &=& g_{(B)}\overline{N}_{(8)} (\sigma^a
- i \gamma_5 \pi^a) {\bf T}^a_{(B)} N_{(9m)} + h.c. \, ,
\end{eqnarray}
together with its mirror part
\begin{eqnarray}\label{eq:lag1}
\mathcal{L}_{(B m)} \nonumber &=& g^\prime_{(B)}\overline{N}_{(8m)}
(\sigma^a + i \gamma_5 \pi^a) {\bf T}^{a}_{(B)} N_{(9)} + h.c. \, .
\end{eqnarray}
At the same time, we have also proven that this is the only
possible case. Moveover, we can easily verify that this Lagrangian
is also invariant under $U_A(1)$ chiral transformation, except
$\mathcal{L}_{(B)}$ and $\mathcal{L}_{(Bm)}$. All these
information is listed in Table~\ref{tab:lags}. Besides these
Lagrangians, we still have the naive combinations: ${m_{(8)}}
\overline{N}_{(8m)} \gamma_5 N_{(8)}$,
${m_{(9)}}\overline{N}_{(9m)} \gamma_5 N_{(9)}$,
${m_{(18)}}\overline{N}_{(18m)} \gamma_5 N_{(18)}$ and
${m_{(10)}}\overline{N}_{(10m)} \gamma_5 N_{(10)}$. There are no
meson fields, but these Lagrangians are still chiral $SU_L(3)
\times SU_R(3)$ invariant and chiral $U(1)_A$ invariant. This
information is listed in Table~\ref{tab:lagsN}.
%===============================================================
\begin{table}[hbt]
\caption{Allowed chiral invariant terms with one meson field. The
$\surd$ denotes that the symmetries are conserved, while $\times$
denotes not.}
\begin{center}\label{tab:lags}
\begin{tabular}{c|c|c|c|c}
\hline \hline ($SU_A(3)$, $U_A(1)$) &
$(\mathbf{1},\mathbf{8})\oplus(\mathbf{8},\mathbf{1})$[mir] &
$(\mathbf{\bar 3},\mathbf{3})\oplus(\mathbf{3},\mathbf{\bar
3})$[mir] & $(\mathbf{6},\mathbf{3}) \oplus (\mathbf{3},\mathbf{6})$
& $(\mathbf{1},\mathbf{10})\oplus(\mathbf{10},\mathbf{1})$[mir]
\\ \hline $(\mathbf{1},\mathbf{8})\oplus(\mathbf{8},\mathbf{1})$[mir] &
N/A & ($\surd$, $\surd$) & ($\surd$, $\surd$) &
N/A \\
\hline $(\mathbf{3},\mathbf{\bar 3})\oplus(\mathbf{\bar
3},\mathbf{3})$[mir] & ($\surd$, $\surd$) & ($\surd$, $\surd$) &
($\surd$, $\surd$) & N/A
\\ \hline $(\mathbf{\bar 6},\mathbf{\bar 3}) \oplus (\mathbf{\bar 3},\mathbf{\bar 6})$
& ($\surd$, $\surd$)
& ($\surd$, $\surd$) & ($\surd$, $\surd$) & ($\surd$, $\surd$)
\\ \hline $(\mathbf{1},\mathbf{\overline{10}})\oplus(\mathbf{\overline{10}},
\mathbf{1})$[mir]& N/A & N/A & ($\surd$, $\surd$)& N/A
\\ \hline
\hline ($SU_A(3)$, $U_A(1)$) &
$(\mathbf{8},\mathbf{1})\oplus(\mathbf{1},\mathbf{8})$ &
$(\mathbf{3},\mathbf{\bar 3})\oplus(\mathbf{\bar 3},\mathbf{3})$
 & $(\mathbf{3},\mathbf{6})\oplus(\mathbf{6},\mathbf{3})$[mir] &
$(\mathbf{10},\mathbf{1}) \oplus (\mathbf{1},\mathbf{10})$
\\ \hline $(\mathbf{8},\mathbf{1})\oplus(\mathbf{1},\mathbf{8})$ &
N/A & ($\surd$, $\surd$) & ($\surd$, $\surd$) &
N/A \\
\hline $(\mathbf{\bar 3},\mathbf{3})\oplus(\mathbf{3},\mathbf{\bar
3})$ & ($\surd$, $\surd$) & ($\surd$, $\surd$) &
($\surd$, $\surd$) & N/A \\
\hline $(\mathbf{\bar 3},\mathbf{\bar 6})\oplus(\mathbf{\bar
6},\mathbf{\bar 3})$[mir] & ($\surd$, $\surd$) & ($\surd$, $\surd$)
& ($\surd$, $\surd$) & ($\surd$, $\surd$)
\\ \hline $(\mathbf{\overline{10}},\mathbf{1}) \oplus (\mathbf{1},\mathbf{\overline{10}})$ &
N/A & N/A & ($\surd$, $\surd$)& N/A
\\ \hline
\hline ($SU_A(3)$, $U_A(1)$) &
$(\mathbf{8},\mathbf{1})\oplus(\mathbf{1},\mathbf{8})$ &
$(\mathbf{1},\mathbf{8})\oplus(\mathbf{8},\mathbf{1})$[mir]
\\ \cline{1-3} $(\mathbf{\bar
3},\mathbf{3})\oplus(\mathbf{3},\mathbf{\bar 3})$ & N/A & ($\surd$, $\times$)  \\
\cline{1-3} $(\mathbf{3},\mathbf{\bar 3})\oplus(\mathbf{\bar
3},\mathbf{3})$[mir] & ($\surd$, $\times$) & N/A
\\ \hline \hline
\end{tabular}
\end{center}
\end{table}
%===============================================================
These results stand in marked contrast to the two-flavor
case~\cite{Dmitrasinovic:2009vp,Dmitrasinovic:2009vy}, where the
$SU_L(2) \times SU_R(2)$ symmetric interactions have both a
$U_A(1)$ symmetry-conserving and a $U_A(1)$ symmetry-breaking
version. Thus, the three-flavor chiral symmetry is more
restrictive than the two-flavor one.
%===============================================================
\begin{table}[hbt]
\caption{Allowed chirally invariant terms without meson field (the
so-called mirror-mass terms). The $\surd$ denotes that the
symmetries are conserved, while $\times$ denotes not.}
\begin{center}\label{tab:lagsN}
\begin{tabular}{c|c|c|c|c}
\hline \hline ($SU_A(3)$, $U_A(1)$) &
$(\mathbf{8},\mathbf{1})\oplus(\mathbf{1},\mathbf{8})$ &
$(\mathbf{3},\mathbf{\bar 3})\oplus(\mathbf{\bar 3},\mathbf{3})$ &
$(\mathbf{3},\mathbf{6})\oplus(\mathbf{6},\mathbf{3})$[mir] &
$(\mathbf{10},\mathbf{1})\oplus(\mathbf{1},\mathbf{10})$
\\ \hline $(\mathbf{1},\mathbf{8})\oplus(\mathbf{8},\mathbf{1})$[mir] &
($\surd$, $\surd$) & N/A & N/A &
N/A \\
\hline $(\mathbf{3},\mathbf{\bar 3})\oplus(\mathbf{\bar
3},\mathbf{3})$[mir] & N/A & ($\surd$, $\surd$) &
N/A & N/A \\
\hline $(\mathbf{\bar 6},\mathbf{\bar 3}) \oplus (\mathbf{\bar
3},\mathbf{\bar 6})$ & N/A & N/A & ($\surd$, $\surd$) & N/A
\\ \hline $(\mathbf{1},\mathbf{\overline{10}})\oplus(\mathbf{\overline{10}},
\mathbf{1})$[mir]& N/A & N/A & N/A & ($\surd$, $\surd$)
\\ \hline\hline
\end{tabular}
\end{center}
\end{table}
%===============================================================

%===============================================================
\section{Chiral mixing}
\label{sect:mix}
%===============================================================

In this section we establish the phenomenologically preferable
mixing pattern(s) and then we use the allowed chiral interactions to
reproduce some of them. First we summarize the salient features of
chiral mixing and axial couplings from Ref.~\cite{Chen:2009sf}.

There are three admissible scenarios (i.e. choices of pairs of
chiral multiplets admixed to the $[(\mathbf{6},\mathbf{3}) \oplus
(\mathbf{3},\mathbf{6})]$ one that lead to real mixing angles) when
fitting the $g_{A}^{(0)}$ and $g_{A}^{(3)}$ that yield the values of
$F$ and $D$. Similarly, when we fit $g_{A}^{(3)}$ and $g_{A}^{(8)}$,
or equivalently $F$ and $D$, we predict the values for $g_{A}^{(0)}$
and $g_{A}^{(3)}$.

This is due to the fact that all three-quark baryon fields satisfy
the relation $g_{A}^{(0)} = 3F - D = \sqrt{3} g_{A}^{(8)}$.
Manifestly, in this way one cannot satisfy both $g_{A~ \rm
expt.}^{(0)} = 0.33 \pm 0.08$ and $g_{A~ \rm expt.}^{(8)} = 0.34
\pm 0.07$. Thus we are left with two possible scenarios:
\begin{enumerate}

\item Fit $g_{A}^{(0)}$ and $g_{A}^{(3)}$ and predict $F$ and $D$. In
Ref.~\cite{Chen:2009sf} we found that there are three possible
mixing patterns. Now the chiral selection rules from
Sect.~\ref{sect:interactions} allow only two of them: the case III-I
mixing (See Table~\ref{tab:spin12b}): $[({\bf 6},{\bf 3})\oplus({\bf
3},{\bf 6})]$--$[(\mathbf{\bar 3},\mathbf{3})\oplus(\mathbf{
3},\mathbf{\bar 3})]$--$[(\mathbf{3},\mathbf{\bar
3})\oplus(\mathbf{\bar 3},\mathbf{3})]$ and the case IV-I mixing:
$[({\bf 6},{\bf 3})\oplus({\bf 3},{\bf
6})]$--$[(\mathbf{1},\mathbf{8})\oplus(\mathbf{8},\mathbf{1})]$--$[(\mathbf{3},\mathbf{\bar
3})\oplus(\mathbf{\bar 3},\mathbf{3})]$. However, the latter mixing
violates $U_A(1)$ symmetry.

\item Fit $g_{A}^{(3)}$ and $g_{A}^{(8)}$ and predict
$g_{A}^{(0)}$. Since we have $g_{A}^{(0)} = \sqrt{3} g_{A}^{(8)}$
for all the three-quark nucleon fields~\cite{Chen:2008qv}, the
results here can be obtained by simple refitting of the previous
case. Fitting $(F,D)$ has not been a problem, so we leave this
exercise out of this paper because it generally overpredicts the
$g_{A}^{(0)}$ by a factor of roughly $\sqrt{3}=1.73$.
\end{enumerate}
Thus we determine the mixing angles in Sect.~\ref{ssect:axcurr},
which we then translate into statements about the admixed fields'
masses in Sect.~\ref{ssect:masses}.
%=============================================
\begin{table}
\caption{The Abelian and the non-Abelian axial charges and the
non-Abelian chiral multiplets of $J^{P}=\frac12$, Lorentz
representation $(\frac{1}{2},0)$ nucleon and $\Delta$ fields, see
Refs.~\cite{Nagata:2007di,Nagata:2008zzc,Dmitrasinovic:2009vp,Dmitrasinovic:2009vy}.}
\label{tab:spin12b}
\begin{tabular}{ccccccccc}
\hline\hline \noalign{\smallskip}
case & field & $g_A^{(0)}$ & $g_A^{(3)}$ & $\sqrt{3} g_A^{(8)}$ &
$F$ & $D$ & $SU_L(3) \times SU_R(3)$ \\
\noalign{\smallskip}\hline\noalign{\smallskip} I & $N_- = N_1 -
N_2$ & $-1$ & $+1$ & $-1$ & $~~0$ & $+1$ &
$(\mathbf{3},\mathbf{\bar
3})\oplus(\mathbf{\bar 3},\mathbf{3})$ \\
II & $N_+ = N_1 + N_2$ & $+3$ & $+1$ & $+3$ & $+1$ & $~~0$ &
$(\mathbf{8},\mathbf{1})\oplus(\mathbf{1},\mathbf{8})$ \\
III & $N_-^\prime$ ($N_-^{(m)}$) & $+1$ & $-1$ & $+1$ & $~~0$ & $-1$
& $(\mathbf{\bar 3},\mathbf{3})\oplus(\mathbf{
3},\mathbf{\bar 3})$ \\
IV & $N_+^\prime$ ($N_+^{(m)}$) & $-3$ & $-1$ & $-3$ & $-1$ & $~~0$ &
$(\mathbf{1},\mathbf{8})\oplus(\mathbf{8},\mathbf{1})$ \\
\hline 0 & $\partial_{\mu} N^\mu$ & $+1$& $+\frac53$ & $+1$ &
$+\frac23$ & $+1$ & $({\bf 6},{\bf 3})\oplus({\bf
3},{\bf 6})$ \\
\noalign{\smallskip}\hline \hline
\end{tabular}
\end{table}
%===========================================
We note here that the relation $g_{A}^{(8)} = \frac{1}{\sqrt{3}}
(3F - D)$ is a general $SU(3)$ result valid for octet fields,
whereas $g_{A}^{(0)} = 3F - D $ is a result that depends on our
specific choice of three-quark interpolating fields being admixed
to the $({\bf 6},{\bf 3})\oplus({\bf 3},{\bf 6})$ one. The latter
relation changes when one considers ``exotic" interpolating
fields, such as certain five-quark (``pentaquark'') ones for
example, and that allows a simultaneous fit of $g_{A}^{(0)},
g_{A}^{(3)}$ and $g_{A}^{(8)}$, which topic is beyond the scope of
this paper.

%===============================================================
\subsection{Phenomenology of the Axial Coupling Constants}
\label{ssect:axcurr}
%===============================================================

A basic feature of the linear chiral realization is that the axial
couplings are determined by the chiral representations. For the
nucleon (proton and neutron), the three-quark chiral representations
of $SU_L(3) \times SU_R(3)$,
$(\mathbf{8},\mathbf{1})\oplus(\mathbf{1},\mathbf{8})$,
$(\mathbf{3},\mathbf{\bar 3})\oplus(\mathbf{\bar 3},\mathbf{3})$ and
$(\mathbf{6},\mathbf{3}) \oplus (\mathbf{3},\mathbf{6})$ provide the
nucleon isovector axial coupling $g_A^{(3)} = 1$, 1 and $5/3$
respectively. Therefore, the mixing of chiral
$(\mathbf{8},\mathbf{1})\oplus(\mathbf{1},\mathbf{8})$,
$(\mathbf{3},\mathbf{\bar 3})\oplus(\mathbf{\bar 3},\mathbf{3})$ and
$(\mathbf{6},\mathbf{3}) \oplus (\mathbf{3},\mathbf{6})$ nucleons
leads to the axial coupling
\begin{eqnarray}
1.267 &=& g_{A~ (\frac12,0)}^{(3)}~\cos^2\theta + g_{A~
(1,\frac12)}^{(3)}~\sin^2\theta
\nonumber\\
&=& g_{A~ (\frac12,0)}^{(3)}~\cos^2\theta + \frac53~\sin^2\theta \,
, \label{e:axcoupl1}
\end{eqnarray}
where $g_{A~ (\frac12,0)}^{(3)}$ represents the coupling of either
$(\mathbf{8},\mathbf{1})\oplus(\mathbf{1},\mathbf{8})$ or
$(\mathbf{3},\mathbf{\bar 3})\oplus(\mathbf{\bar 3},\mathbf{3})$,
and $g_{A~ (1,\frac12)}^{(3)}$ represents the coupling of
$(\mathbf{6},\mathbf{3}) \oplus (\mathbf{3},\mathbf{6})$. The
coupling $g_{A~ (1,\frac12)}^{(3)}$ is needed because only the
coupling of $(\mathbf{6},\mathbf{3}) \oplus (\mathbf{3},\mathbf{6})$
is larger than the experimental value $1.267$. We list the results
of the mixing angles for all the four cases in
Table~\ref{tab:axcoupl1r}.
%=============================================
\begin{table}[tbh]
\begin{center}
\caption{The values of the baryon isoscalar axial coupling constant
predicted from the naive mixing and $g_{A~ \rm expt.}^{(3)}=1.267$;
compare with $g_{A~ \rm expt.}^{(0)}=0.33 \pm 0.03 \pm 0.05$,
$F$=$0.459\pm0.008$ and $D$=$0.798\pm0.008$, leading to $F/D = 0.571
\pm 0.005$, Ref.~\cite{Yamanishi:2007zza}.}
\begin{tabular}{ccccccccc}
\hline \hline case & $g_{A~ \rm expt.}^{(3)}$ & $\theta_{i}$ &
$g_{A~\rm mix.}^{(0)}$ & $\sqrt{3} g_{A~\rm mix.}^{(8)}$ & $F$ & $D$ & $F$/$D$ \\
\hline
I & 1.267 & $39.3^o$ & $-0.20$ & $-0.20$ & 0.267 & 1 & 0.267 \\
II & 1.267 & $39.3^o$ & 2.20 & 2.20 & 0.866 & 0.401 & 2.16 \\
III & 1.267 & $67.2^o$ & 1.00 & 1.00 & 0.567 & 0.700 &  0.81  \\
IV & 1.267 & $67.2^o$ & 0.40 & 0.40 & 0.417 & 0.850 & 0.491 \\
 \hline \hline
\end{tabular}
\label{tab:axcoupl1r}
\end{center}
\end{table}
%=============================================
Three-quark nucleon interpolating fields in QCD have well-defined
$U_A(1)$ chiral transformation properties, see Table
\ref{tab:spin12b}, that can be used to predict the flavor singlet
axial coupling $g_{A~\rm mix.}^{(0)}$ and the $F$ and $D$ values
\begin{eqnarray}
g_{A~\rm mix.}^{(0)} &=& g_{A~ (\frac12,0)}^{(0)}~\cos^2\theta +
g_{A~ (1,\frac12)}^{(0)}~\sin^2\theta
\nonumber \\
&=& g_{A~ (\frac12,0)}^{(0)}~\cos^2\theta + \sin^2\theta ,
\label{e:axcoupl0} \\
F &=& F_{(\frac12,0)}~\cos^2\theta +
F_{(1,\frac12)}^{(1)}~\sin^2\theta ,
\nonumber\\
&=& F_{(\frac12,0)}~\cos^2\theta + \frac23~\sin^2\theta \label{e:F} \, , \\
D &=& D_{(\frac12,0)}~\cos^2\theta + D_{(1,\frac12)}~\sin^2\theta
\nonumber \\
&=& D_{(\frac12,0)}~\cos^2\theta + \sin^2\theta \, . \label{e:D}
\end{eqnarray}
The mixing angle $\theta$ is extracted from Eq.~(\ref{e:axcoupl1}),
where we used the bare $F$ and $D$ values for different chiral
multiplets as listed in Table~\ref{tab:spin12b}. Due to the
different (bare) non-Abelian $g_{A}^{(3)}$ and Abelian $g_{A}^{(0)}$
axial couplings, see Table~\ref{tab:spin12b}, the mixing formulae
Eq.~(\ref{e:axcoupl0}) give substantially different predictions from
one case to another, see Table~\ref{tab:axcoupl1r}. We can see in
Table~\ref{tab:axcoupl1r} that the two best candidates are cases I
and IV, with $g_A^{(0)} = - 0.2$ and $g_{A}^{(0)} = 0.4$,
respectively, the latter being within the error bars of the measured
value $g_{A~ \rm expt.}^{(0)} = 0.33 \pm
0.08$~\cite{Bass:2007zzb,Ageev:2007du}. Selection rules from
Sect.~\ref{sect:interactions} allow the case III and the case IV.
And so the case IV is the best candidate so long as we consider just
the mixing of two nucleon fields~\cite{Dmitrasinovic:2009vp}.

Manifestly, a linear superposition of any three fields (except for
the mixtures of cases II and III, IV above, which yield complex
mixing angles) gives a perfect fit to the central values of the
experimental axial couplings $g_{A~ \rm expt.}^{(0)} = 0.33 \pm
0.08$ and $g_{A~ \rm expt.}^{(3)}=1.267$ and predict the $F$ and
$D$ values, or {\it vice versa}: one may fit $g_A^{(3)}$ and
$g_A^{(8)}$, (or equivalently $F$ and $D$) and thus predict
$g_A^{(0)}$. This has been done in Ref.~\cite{Chen:2009sf}, and
where there were three allowed cases: I-II, I-III and I-IV. The
selection rules from Sect.~\ref{sect:interactions} indicate that
only two of them are possible in the one-meson approximation: (1)
the case I-III and (2) the case I-IV. In the former case the
$U_A(1)$ symmetry is conserved, whereas in the latter the $U_A(1)$
is violated.

Such a three-field admixture introduces two new free parameters,
besides the already introduced mixing angles, e.g. $\theta_{3}$
and $\theta_{1}$(=0), (which we may set to vanish in the present
approximation). For the case I-III (we shall call it here case
III-I for reasons soon to be clarified) we have the
relative/mutual mixing angle $\theta_{31} = \varphi$, as the two
nucleon fields III and I mix due to the off-diagonal interaction
Eq.~(\ref{e:8/9a interact}). Thus we find two equations with two
unknowns of the general form:
\begin{align}
\frac{5}{3}\,{\sin}^2 \theta + {\cos}^2\theta\,\left(g_A^{(3)}({\rm
III})
{\cos}^2 \varphi + g_A^{(3)}({\rm I}){\sin}^2\varphi \right) &= 1.267 \, , \\
{\sin}^2 \theta + {\cos}^2 \theta\, \left(g_A^{(0)}({\rm
III}){\cos}^2 \varphi + g_A^{(0)}({\rm I})\,{\sin}^2 \varphi \right)
&= 0.33 \pm 0.08 \, .
\end{align}
The solutions to these equations (the values of the mixing angles
$\theta,\varphi$) provide, at the same time, input for the
prediction of $F$ and $D$:
\begin{align}
\cos^2\theta\,\left(F({\rm III})\, {\cos}^2 \varphi + F({\rm I})
\,{\sin}^2\varphi \right) + \frac23~\sin^2\theta
&= F  \label{e:F1} \, , \\
\cos^2\theta\,\left(D({\rm III})\, {\cos}^2 \varphi + D({\rm I})
\,{\sin}^2\varphi \right) + \sin^2\theta &= D \, . \label{e:D1}
\end{align}
The values of the mixing angles ($\theta,\varphi$) obtained from
this straightforward fit to the baryon axial coupling constants are
shown in Table~\ref{tab:axcoupl2b}. We also show the result of the
case I-IV as well as IV-I in this Table. Besides these cases, the
cases I-II and II-I can also be used to produce the experimental
$g_A^{(0)}$ and $g_A^{(3)}$, which are however not allowed from
Sect~\ref{sect:interactions}.

%=============================================
\begin{table}[tbh]
\begin{center}
\caption{The values of the mixing angles obtained from the simple
fit to the baryon axial coupling constants and the predicted
values of axial $F$ and $D$ couplings. The experimental values are
$F$=$0.459\pm0.008$ and $D$=$0.798\pm0.008$, leading to $F/D =
0.575 \pm 0.005$ and $g_{A}^{(8)} = 0.33 \pm 0.01$,
Ref.~\cite{Ratcliffe:1990dh}. The most recent analysis of
experimental values leads to $F=0.477 \pm 0.001$ and $D=0.835 \pm
0.001$ and $g_{A}^{(8)} = 0.344 \pm 0.001$ in
Ref.~\cite{Yamanishi:2007zza}. Note that these values are more
than 2-$\sigma$ away from the old ones, and that the new $F$,$D$
add up to $F+D$ = 1.312 $\neq 1.269 \pm 0.002$. Also $g_{A~ \rm
expt.}^{(0)} = 0.33 \pm 0.08$.}
\begin{tabular}{ccccccccc}
\hline \hline case & $g_{A~ \rm expt.}^{(3)}$ & $g_{A}^{(0)}$ &
$g_{A}^{(8)}$ &
$\theta$ & $\varphi$ & $F$ & $D$ & $F$/$D$ \\
\hline I-III & 1.267 & $0.33 \pm 0.08$ & $0.19 \pm 0.05$ & $50.7^o
\pm 1.8^o$ & $23.9^o \pm 2.9^o$ & $0.399 \pm 0.02$ & 0.868$\mp
0.02$ &
$0.460 \pm 0.04$ \\
III-I & 1.267 & $0.33 \pm 0.08$ & $0.19 \pm 0.05$ & $50.7^o \pm
1.8^o$ & $66.1^o \pm 2.9^o$ & $0.399 \pm 0.02$ & 0.868$\mp 0.02$ &
$0.460 \pm 0.04$ \\
\hline I-IV & 1.267 & $0.33 \pm 0.08$ & $0.19 \pm 0.05$ & $63.2^o
\pm 4.0^o$ & $54^o \pm 23^o$ & $0.399 \pm 0.02$ & 0.868$\mp 0.02$
&
$0.460 \pm 0.04$ \\
IV-I & 1.267 & $0.33 \pm 0.08$ & $0.19 \pm 0.05$ & $63.2^o \pm
4.0^o$ & $36^o \pm 23^o$ & $0.399 \pm 0.02$ & 0.868$\mp 0.02$ &
$0.460 \pm 0.04$ \\
\hline \hline
\end{tabular}
\label{tab:axcoupl2b}
\end{center}
\end{table}
%=============================================

\subsection{Baryon Masses} \label{ssect:masses}

The next step is to try and reproduce this phenomenological mixing
starting from a model interaction, rather than {\it per fiat}. As
the first step in that direction we must look for a dynamical source
of mixing. One such mechanism is the simplest chirally symmetric
{\it non-derivative} one-$(\sigma,\pi)$-meson interaction
Lagrangian, which induces baryon masses via its $\sigma$-meson
coupling. Chiral symmetry is spontaneously broken through the
``condensation'' of the sigma field $\sigma \rightarrow
\sigma_0=\langle \sigma \rangle_0 = f_\pi$, which leads to the
dynamical generation of baryon masses, as can be seen from the
linearized chiral invariant interaction Lagrangians
Eqs.~(\ref{def:18 int}) and (\ref{def:9 int}).

In this section, we study the masses of the octet baryons. There are
altogether six types of octet baryon fields: $N_+$ ($N_{(8)}$),
$N_-$ (contained in $N_{(9)}$) and $N_\mu$ (contained in
$N_{(18)}$), as well as their mirror fields $N_+^{\prime}$
($N_{(8m)}$), $N_-^{\prime}$ (contained in $N_{(9m)}$),
$N^\prime_\mu$ (contained in $N_{(18m)}$). The nucleon mass matrix
is already in a simple block-diagonal form when the nucleon fields
form the following mass matrix:
\begin{eqnarray}
M &=& {1\over\sqrt{6}}\bar N \left ( \begin{array}{ccc|ccc} 0 &
g_{(8/9)} & g_{(8/18)} & m_{(8)} \gamma_5 & g_{B} & 0
\\ g^*_{(8/9)} &
g_{(9/9)} & g_{(9/18)} & g^*_{B} & m_{(9)} \gamma_5 & 0
\\ g^*_{(8/18)} &
g^*_{(9/18)}& g_{(18/18)} & 0 & 0 & m_{(18)} \gamma_5
\\ \hline
m_{(8)} \gamma_5 & g^\prime_{B} & 0 & 0 & g^\prime_{(8/9)} &
g^\prime_{(8/18)}
\\ g^{\prime*}_{B} & m_{(9)} \gamma_5 & 0 & g^{\prime*}_{(8/9)} &
g^\prime_{(9/9)} & g^\prime_{(9/18)}
\\ 0 & 0 & m_{(18)} \gamma_5 & g^{\prime*}_{(8/18)} &
g^{\prime*}_{(9/18)}& g^\prime_{(18/18)}
\end{array} \right ) N \, ,
\end{eqnarray}
where
\begin{eqnarray}
N = ( N_+^{\prime}, N_-^{\prime}, N_\mu, N_+, N_-, N_\mu^\prime )^T
\, .
\end{eqnarray}
Since there are three nucleon fields as well as their mirror fields,
there can be a nonzero phase angle. However, for simplicity, we
assume all the axial couplings are real.

%===============================================================
\subsection{Masses due to $[({\bf 6},{\bf 3})\oplus({\bf
3},{\bf 6})]$--$[(\mathbf{\bar 3},\mathbf{3})\oplus(\mathbf{
3},\mathbf{\bar 3})]$ mixing} \label{ssect:(6,3)(3,3)masses}
%===============================================================

We use the results of Sect.~\ref{sect:interactions}: the chirally
invariant diagonal, Eqs.~(\ref{def:18 int}) and (\ref{def:9 int})
and off-diagonal, Eq.~(\ref{e:9/18 interact}) meson-baryon-baryon
interactions involving
\begin{eqnarray}
\nonumber (B_1,\Lambda) &\in& (\mathbf{\bar
3},\mathbf{3})\oplus(\mathbf{ 3},\mathbf{\bar 3}) [{\rm mir}] \, ,
\\ (B_2,\Delta) &\in& ({\bf 6},{\bf 3})\oplus({\bf 3},{\bf 6}) \, ,
\\ \nonumber (\sigma,\pi) &\in& (\mathbf{\bar 3},\mathbf{3})\oplus(\mathbf{
3},\mathbf{\bar 3}) \, .
\end{eqnarray}
Here all baryons have spin 1/2, while the isospin of $B_1$ and $B_2$
is 1/2 and that of $\Delta$ is 3/2. The $\Delta$ field is then
represented by an isovector-Diracspinor field $\Delta^i$,
($i=1,2,3$).

In writing down the Lagrangians Eqs.~(\ref{def:18 int}),
(\ref{def:9 int}) and (\ref{e:9/18 interact}), we have implicitly
assumed that the parities of $B_1$, $B_2$, $\Lambda$ and $\Delta$
are the same. In principle, they are arbitrary, except for the
ground state nucleon, which must be even. For instance, if $B_2$
has odd parity, the first term in the interaction Lagrangian
Eq.~(\ref{e:9/18 interact}) must include another $\gamma_5$
matrix~\cite{Jido:2001nt}. Here we assume the ground state nucleon
is contained in either $[({\bf 6},{\bf 3})\oplus({\bf 3},{\bf
6})]$ or $[(\mathbf{\bar 3},\mathbf{3})\oplus(\mathbf{
3},\mathbf{\bar 3})]$, and so at least one of $B_1$ and $B_2$ has
even parity. Next we consider all possible cases for the parities
of $B_2$, $\Lambda$ and $\Delta$. The results are similar to the
two-flavor ones shown in
Ref.~\cite{Dmitrasinovic:2009vp,Dmitrasinovic:2009vy} (because we
assumed good $SU(3)$ symmetry here).

Having established the mixing interaction Eq.~(\ref{e:9/18
interact}), as well as the diagonal terms Eqs.~(\ref{def:18 int})
and (\ref{def:9 int}), we calculate the masses of the baryon states,
as functions of the pion decay constant/chiral order parameter and
the coupling constants $g_1\sim g_{(9)}$, $g_2\sim g_{(18)}$ and
$g_3\sim g_{(9/18)}$:
\begin{eqnarray}\label{def:33}
\nonumber \mathcal{L}_{(9)} &=& - g_1 \Big ( \bar B_1 \sigma B_1 - 2
\bar \Lambda \sigma \Lambda \Big ) + \cdots \, ,
\\ \mathcal{L}_{(18)} &=& - g_2 \Big ( \bar B_2 \sigma B_2 -
2 \bar \Delta^i \sigma \Delta^i \Big ) + \cdots \, ,
\\ \nonumber \mathcal{L}_{(9/18)} &=& - g_3 \Big ( \bar B_1 \sigma B_2 \Big ) + \cdots \, ,
\end{eqnarray}
Altogether we have
\begin{eqnarray}
\mathcal{L} &=& - f_\pi (\bar B_1, \bar B_2) \left(
\begin{array}{cc} g_1 & g_3
\\ g_3 & g_2
\end{array} \right ) \left(
\begin{array}{c} B_1
\\ B_2
\end{array} \right ) + 2 g_1 f_\pi \bar \Lambda \Lambda + 2 g_2 f_\pi \bar \Delta^i \Delta^i
\end{eqnarray}
We diagonalize the mass matrix and express the mixing angle in terms
of diagonalized masses
\begin{eqnarray}
N(N^*) &=& \cos \theta B_1 + \sin \theta B_2 \, ,
\\ \nonumber N^*(N) &=& - \sin \theta B_1 + \cos \theta B_2 \, .
\end{eqnarray}
We find the following double-angle formulas for the mixing angles
$\theta_{1,\cdots,8}$ between $B_1$ and $B_2$ in the eight different
parities scenarios
\begin{eqnarray}
\tan 2\theta_1 &=& \frac{\sqrt{-(2 N + \Delta )(2 N^{*} +
\Delta})}{(N + N^{*} + \Delta)} = - \frac{\sqrt{-(2 N + \Lambda )(2
N^{*} + \Lambda})}{(N + N^{*} + \Lambda)} , \label{e:mix1a}
\\ \tan 2\theta_2 &=& \frac{\sqrt{(2 N + \Delta )(2 N^{*} -
\Delta})}{(N - N^{*} + \Delta)} = - \frac{\sqrt{(2 N + \Lambda )(2
N^{*} - \Lambda})}{(N - N^{*} + \Lambda)} , \label{e:mix2a} \\
\tan 2\theta_3 &=& \frac{\sqrt{-(2 N + \Delta )(2 N^{*} +
\Delta})}{(N + N^{*} + \Delta)} = - \frac{\sqrt{-(2 N - \Lambda )(2
N^{*} - \Lambda})}{(N + N^{*} - \Lambda)} , \label{e:mix3a} \\
\tan 2\theta_4 &=& \frac{\sqrt{(2 N + \Delta )(2 N^{*} -\Delta})}{(N
- N^{*} + \Delta)} =  \frac{\sqrt{(2 N - \Lambda )(2
N^{*} + \Lambda})}{(-N + N^{*} + \Lambda)} , \label{e:mix4a} \\
\tan 2\theta_5 &=& \frac{\sqrt{-(2 N - \Delta )(2 N^{*}
-\Delta})}{(N + N^{*} - \Delta)} = - \frac{\sqrt{-(2 N + \Lambda )(2
N^{*} + \Lambda})}{(N + N^{*} + \Lambda)} , \label{e:mix5a} \\
\tan 2\theta_6 &=& \frac{\sqrt{(2 N - \Delta )(2 N^{*} +
\Delta})}{(N - N^{*} - \Delta)} = - \frac{\sqrt{(2 N + \Lambda )(2
N^{*} - \Lambda})}{(N - N^{*} + \Lambda)} , \label{e:mix6a} \\
\tan 2\theta_7 &=&  \frac{\sqrt{-(2 N - \Delta )(2 N^{*}
-\Delta})}{(N + N^{*} - \Delta)} = - \frac{\sqrt{-(2 N - \Lambda )(2
N^{*} - \Lambda})}{(N + N^{*} - \Lambda)} , \label{e:mix7a} \\
\tan 2\theta_8 &=& \frac{\sqrt{(2 N - \Delta )(2 N^{*} +
\Delta})}{(N - N^{*} - \Delta)} = \frac{\sqrt{(2 N - \Lambda )(2
N^{*} + \Lambda})}{(N - N^{*} - \Lambda)} , \label{e:mix8a}
\end{eqnarray}
where $N$, $N^{*}$, $\Lambda$ and $\Delta$ represent the masses of
the corresponding particles. The four angles correspond to the eight
possible parities; $\theta_1: (N^{*+},\Lambda^+,\Delta^{+})$,
$\theta_2: (N^{*-},\Lambda^+,\Delta^{+})$, $\theta_3:
(N^{*+},\Lambda^-,\Delta^{+})$, $\theta_4:
(N^{*-},\Lambda^-,\Delta^{+})$, $\theta_5:
(N^{*+},\Lambda^+,\Delta^{-})$, $\theta_6:
(N^{*-},\Lambda^+,\Delta^{-})$, $\theta_7:
(N^{*+},\Lambda^-,\Delta^{-})$, $\theta_8:
(N^{*-},\Lambda^-,\Delta^{-})$, where $\pm$ indicate the parity of
the state. Note that the angle $\theta_1$, $\theta_3$ and $\theta_5$
is necessarily imaginary so long as the $\Delta$, $\Lambda$ and
$N^{*}$ masses are physical (positive), and that the reality of the
mixing angle(s) imposes stringent limits on the $\Delta, N^{*}$
resonance masses in other cases, as well.

In the present study we have three model parameters $g_1, g_2$ and
$g_3$, which can be determined by different set of inputs. We can
use two baryon masses and the mixing angle as inputs and predicts
the third baryon mass (Inverse prediction). We use the formulas
Eqs.~(\ref{e:mix1a})-(\ref{e:mix8a}) for the (double) mixing angles
$\theta_{1,...,8}$ together with the two observed nucleon masses and
the mixing angle $\theta = 67.2^o$ as shown in
Table~\ref{tab:axcoupl1r} to predict the $\Delta$ masses shown in
the Table~\ref{tab:prediction1}.

%=============================================
\begin{table}[tbh]
\begin{center}
\caption{The values of the $\Delta$ baryon masses predicted from the
isovector axial coupling $g_{A~\rm mix.}^{(1)} = g_{A~ \rm
expt.}^{(1)} = 1.267$ and $g_{A~\rm mix.}^{(0)}=0.4$ vs. $g_{A~ \rm
expt.}^{(0)} = 0.33 \pm 0.08$.}
\begin{tabular}{c|c|cc|cc}
\hline \hline $(N^{*P},\Lambda^{P^\prime},\Delta^{P^{''}})$ & ($N,
~~N^{*}$) & $\Lambda$ (MeV) & $\Lambda_{\rm expt.}$ (MeV) & $\Delta$
(MeV) & $\Delta_{\rm expt.}$ (MeV)  \\
\hline $(-,+,+)$ & N(940), R(1535) & 2330 & - & 2330 & 1910 \\
$(-,-,+)$ & N(940), R(1535) & 1140 & 1405 & 2330 & 1910 \\
$(-,+,-)$ & N(940), R(1535) & 2330 & - & 1140 & - \\
$(+,-,-)$ & N(940), R(1440) & 2030,2730 & - & 2030,2730 & - \\
$(-,-,-)$ & N(940), R(1535) & 1140 & 1405 & 1140 & - \\
 \hline \hline
\end{tabular}
\label{tab:prediction1}
\end{center}
\end{table}
%=============================================

We see that only the $(N^{*-},\Delta^{+})$ parity combination
leads to a realistic prediction of the baryon masses. Otherwise,
at least one of the predicted baryon masses is off by a factor of
of order two. Indeed, the case
$(N^{*P},\Lambda^{P^\prime},\Delta^{P^{''}}) = (-,-,+)$ predicts
the (odd-parity) $SU(3)$ flavor-singlet $\Lambda$ at 1140 MeV,
somewhat below the measured value (1405 MeV) and $\Delta(2330)$,
the nearest known candidate state being the (four star PDG, Ref.
\cite{Amsler:2008zzb}) $P_{31}(1910)$ resonance. It is curious
that the flavor-singlet $\Lambda(1140)$ state lies (considerably)
below the flavor-octet state $N^{*}(1535)$ even in the good flavor
$SU(3)$ symmetry limit; the predicted mass difference might/ought
to be improved by introducing explicit $SU(3)$ symmetry breaking
strange-up/down quark mass difference.

%===============================================================
\subsection{Masses due to $[({\bf 6},{\bf 3})\oplus({\bf
3},{\bf 6})]$--$[(\mathbf{\bar 3},\mathbf{3})\oplus(\mathbf{
3},\mathbf{\bar 3})]$--$[(\mathbf{3},\mathbf{\bar 3})\oplus(\mathbf{
\bar 3},\mathbf{3})]$ mixing} \label{ssect:(6,3)(3,3)masses}
%===============================================================

To improve our analysis, we may add a third chiral multiplet nucleon
field. As in the previous section~\ref{sect:interactions}, we
consider baryon fields
\begin{eqnarray}
\nonumber (B_1,\Lambda_1) &\in& (\mathbf{\bar
3},\mathbf{3})\oplus(\mathbf{ 3},\mathbf{\bar 3}) [{\rm mir}] \, ,
\\ (B_2,\Delta) &\in& ({\bf 6},{\bf 3})\oplus({\bf 3},{\bf 6}) \, ,
\\ \nonumber (B_3,\Lambda_2) &\in& (\mathbf{3},\mathbf{\bar
3})\oplus(\mathbf{\bar 3},\mathbf{ 3}) \, .
\end{eqnarray}
As discussed above, the case III-I allows one to reproduce the
experimental couplings $g_A^{(0)}$ and $g_A^{(3)}$. To study this
mixing, we need to use the previous Lagrangian Eq. (\ref{def:33}) as
well as the new ones:
\begin{eqnarray}
\label{def:33_2} \nonumber \mathcal{L}^\prime_{(9)} &=& - g_4 \Big (
\bar B_3 \sigma B_3 - 2 \bar \Lambda_1 \sigma \Lambda_1 \Big ) + \cdots \, ,\\
\mathcal{L}_{(9/9)} &=& - g_5 f_\pi \bar B_1 B_3 - g_5 f_\pi \bar
\Lambda_1 \Lambda_2 + \cdots \, ,
\end{eqnarray}
that follow from Eq. (\ref{def:9 int}), where the the third nucleon
field $B_3$ is a mirror image of $B_1$. We note that $B_1$ and $B_3$
couple with each other through the naive combinations:
${m_{(9)}}\overline{N}_{(9m)} \gamma_5 N_{(9)}$. Chiral symmetry is
spontaneously broken through the ``condensation'' of the sigma field
$\sigma \to \sigma_0 = \langle \sigma \rangle_{0} = f_{\pi}$, which
leads to the dynamical generation of baryon masses:
\begin{eqnarray}
\mathcal{L} &=& - f_\pi (\bar B_1, \bar B_3 , \bar B_2) \left(
\begin{array}{ccc} g_1 & g_5 & g_3
\\ g_5 & g_4 & 0
\\ g_3 & 0 & g_2
\end{array} \right ) \left(
\begin{array}{c} B_1
\\ B_3
\\ B_2
\end{array} \right ) - f_\pi (\bar \Lambda_1 , \bar \Lambda_2) \left(
\begin{array}{cc} - 2 g_1 & g_5
\\ g_5 & - 2 g_4
\end{array} \right ) \left(
\begin{array}{c} \Lambda_1
\\ \Lambda_2
\end{array} \right ) + 2 g_2 f_\pi \bar \Delta^i \Delta^i \nonumber
\\
\end{eqnarray}
To solve this system in its full generality seems both too
complicated and not very useful. However, since $g_6$ of $g_6 \bar
B_3 B_2$ vanishes, we only need five conditions to solve this
system. Therefore, we just use the three nucleon candidates
$N(940)$, $N(1440)$ and $N^*(1535)$ as well as the two mixing
angles $\theta ^o= 63.2^o$ and $\phi = 36^o$. Finally we find that
there are two possibilities as shown in
Table~\ref{tab:prediction4}.

%=============================================
\begin{table}[tbh]
\begin{center}
\caption{The values of the $\Delta$ and $\Lambda$ baryon masses
predicted from the isovector axial coupling $g_{A~\rm mix.}^{(1)}
= g_{A~ \rm expt.}^{(1)} = 1.267$ and $g_{A~\rm mix.}^{(0)} = 0.33
\pm 0.08$ due to $[({\bf 6},{\bf 3})\oplus({\bf 3},{\bf 6})]$ --
$[(\mathbf{\bar 3},\mathbf{3})\oplus(\mathbf{ 3},\mathbf{\bar
3})]$ -- $[(\mathbf{3},\mathbf{\bar 3})\oplus(\mathbf{ \bar
3},\mathbf{3})]$ mixing .}
\begin{tabular}{c|ccccc|ccc}
\hline \hline No. & $g_1$ & $g_2$ & $g_3$ & $g_4$ & $g_5$ &
$\Lambda_1^P$ (MeV) & $\Lambda_2^P$ (MeV) & $\Delta^P$ (MeV) \\
\hline 1 & $-4.7$ & 8.4 & $-3.4$ & 2.9 & 9.8 & $1370^-$ & $1850^+$ &
$2170^-$
\\ 2 & $-7.2$ & 4.6 & 7.9 & 9.1 & $-4.2$ & $1940^+$ & $2430^-$ & $1200^-$
\\ \hline \hline
\end{tabular}
\label{tab:prediction4}
\end{center}
\end{table}
%=============================================

Once again, the odd-parity $\Delta$ option appears as the better
one. Now, the first flavor-singlet $\Lambda$ lies at 1370 MeV,
substantially closer to 1405 MeV than before. A second
flavor-singlet $\Lambda$ lies at 1850 MeV, very close to the
(three star PDG, Ref. \cite{Amsler:2008zzb}) $P_{01}(1810)$
resonance. This is our best candidate in the $[({\bf 6},{\bf
3})\oplus({\bf 3},{\bf 6})]$--$[(\mathbf{\bar
3},\mathbf{3})\oplus(\mathbf{ 3},\mathbf{\bar
3})]$--$[(\mathbf{3},\mathbf{\bar 3})\oplus(\mathbf{ \bar
3},\mathbf{3})]$ mixing scenario.
%Yet another solution is not out of the question, however: the first
%flavor-singlet $\Lambda$ at 1940 MeV is not far from the (four
%star PDG) $P_{03}(1890)$ resonance. Yet, the second flavor-singlet
%$\Lambda$ lies at 2430 MeV cannot be found in PDG tables.

%===============================================================
\subsection{Masses due to $[({\bf 6},{\bf 3})\oplus({\bf
3},{\bf 6})]$--$[(\mathbf{1},\mathbf{8})\oplus(\mathbf{
8},\mathbf{1})]$ mixing} \label{ssect:(6,3)(1,8)masses}
%===============================================================

We can also study the baryon masses due to $[({\bf 6},{\bf
3})\oplus({\bf 3},{\bf 6})]$ -- $[(\mathbf{1},\mathbf{8}) \oplus
(\mathbf{ 8},\mathbf{1})]$ mixing
\begin{eqnarray}
\nonumber B_1 &\in& (\mathbf{1},\mathbf{8}) \oplus (\mathbf{
8},\mathbf{1}) [{\rm mir}] \, ,
\\ (B_2,\Delta) &\in& ({\bf 6},{\bf 3})\oplus({\bf 3},{\bf 6}) \, .
\end{eqnarray}
Having established the mixing interaction Eq.~(\ref{e:8/18
interact}), as well as the diagonal terms Eq.~(\ref{def:18 int}), we
calculate the masses of the baryon states, as functions of the pion
decay constant/chiral order parameter and the coupling constants
$g_2\sim g_{(18)}$ and $g_3\sim g_{(8/18)}$:
\begin{eqnarray}
\label{def:18} \mathcal{L}_{(18)} &=& - g_2 \Big( \bar B_2 \sigma
B_2 - 2 \bar \Delta^i \sigma \Delta^i \Big) + \cdots \, ,
\\ \nonumber \mathcal{L}_{(8/18)} &=& - g_3 \Big( \bar B_1 \sigma B_2 \Big) + \cdots \, ,
\end{eqnarray}
Note that $g_1\sim g_{(8)}$ is zero now. We diagonalize the mass
matrix and express the mixing angle in terms of diagonalized masses.
We find the following double-angle formulas for the mixing angles
$\theta_{1,\cdots,4}$ between $B_1$ and $B_2$ in the four different
parities scenarios
\begin{eqnarray}
\tan 2\theta_1 &=& -2i \frac{\sqrt{N N^{*}} }{N^{*} + N } \, ,
\Delta
= -2 (N^* + N ) \label{e:mix1b} \, , \\
\tan 2\theta_2 &=&  \frac{2\sqrt{N N^{*}} }{ N^{*} - N} \, , \Delta
= 2 (N^* - N ) \label{e:mix2b} \, , \\
\tan 2\theta_3 &=& -2i \frac{\sqrt{N N^{*}} }{N^{*} + N } \, ,
\Delta = 2 (N^* + N ) \label{e:mix3b} \, , \\
\tan 2\theta_4 &=&  \frac{2\sqrt{N N^{*}} }{ N^{*} - N} \, ,
\Delta = -2 (N^* - N ) \label{e:mix4b} \, ,
\end{eqnarray}
where $N$, $N^{*}$ and $\Delta$ represent the masses of the
corresponding particles. The four angles correspond to the four
possible parities; $\theta_1: (N^{*+},\Delta^{+})$, $\theta_2:
(N^{*-},\Delta^{+})$, $\theta_3: (N^{*+},\Delta^{-})$, $\theta_4:
(N^{*-},\Delta^{-})$, where $\pm$ indicate the parity of the
state. Note that only $\theta_2$ leads to a physical result. We
can use the mixing angle $\theta = 67.2^o$ and the nucleon mass
940 MeV to predict the excited nucleon mass and $\Delta$ mass, see
Table \ref{tab:prediction2a}.
%=============================================
\begin{table}[!tbh]
\begin{center}
\caption{The values of the $\Delta$ baryon masses predicted from
the isovector axial coupling $g_{A~\rm mix.}^{(1)} = g_{A~ \rm
expt.}^{(1)} = 1.267$ and $g_{A~\rm mix.}^{(0)}=0.4$ vs. $g_{A~
\rm expt.}^{(0)} = 0.33 \pm 0.08$ due to $[({\bf 6},{\bf
3})\oplus({\bf 3},{\bf 6})]$ -- $[(\mathbf{1},\mathbf{8}) \oplus
(\mathbf{ 8},\mathbf{1})]$ mixing without additional two-meson
interactions.}
\begin{tabular}{c|c|cc|cc}
\hline \hline $(N^{*P},\Delta^{P^{'}})$ & $N$ & $N^{*}$ & $N^*_{\rm
expt.}$ (MeV) & $\Delta$
(MeV) & $\Delta_{\rm expt.}$ (MeV)  \\
\hline $(-,+)$ & N(940) & 5320 & - & 8760 & - \\
\hline \hline
\end{tabular}
\label{tab:prediction2a}
\end{center}
\end{table}
%=============================================
This gives predictions of no practical value. To get a practically
useful result, we need to add one of the two-meson interaction
Lagrangians from Sect. \ref{ssect:(8,1)Baryons int}, and thus a
non-zero $g_1$ term:
\begin{eqnarray}
\nonumber \mathcal{L}_{(8)} &=& - {g_1\over f_\pi} \bar B_1 \sigma^2
B_1 + \cdots \, ,
\end{eqnarray}
and we have four new different parities scenarios:
\begin{eqnarray}
\tan 2\theta_1 &=& \frac{\sqrt{-(2 N + \Delta )(2 N^{*} +
\Delta})}{(N + N^{*} + \Delta)}\, , \label{e:mix1c}
\\ \tan 2\theta_2 &=& \frac{\sqrt{(2 N + \Delta )(2 N^{*} -
\Delta})}{(N - N^{*} + \Delta)} \, , \label{e:mix2c} \\
\tan 2\theta_3 &=& \frac{\sqrt{-(2 N - \Delta )(2 N^{*}
-\Delta})}{(N + N^{*} - \Delta)} \, , \label{e:mix3c} \\
\tan 2\theta_4 &=& \frac{\sqrt{(2 N - \Delta )(2 N^{*} +
\Delta})}{(N - N^{*} - \Delta)} \, . \label{e:mix4c}
\end{eqnarray}
Note that only $\theta_1$ is imaginary for positive baryon masses,
i.e. unphysical. We can use the mixing angle $\theta = 67.2^o$ and
the two nucleon masses to predict the $\Delta$ mass, see Table
\ref{tab:prediction3a}. The nearest known candidate for the
$\Delta(2330)$ state is the (four star PDG, Ref.
\cite{Amsler:2008zzb}) $P_{31}(1910)$ resonance.
%=============================================
\begin{table}[tbh]
\begin{center}
\caption{The values of the $\Delta$ baryon masses predicted from
the isovector axial coupling $g_{A~\rm mix.}^{(1)} = g_{A~ \rm
expt.}^{(1)} = 1.267$ and $g_{A~\rm mix.}^{(0)}=0.4$ vs. $g_{A~
\rm expt.}^{(0)} = 0.33 \pm 0.08$ due to $[({\bf 6},{\bf
3})\oplus({\bf 3},{\bf 6})]$ -- $[(\mathbf{1},\mathbf{8}) \oplus
(\mathbf{ 8},\mathbf{1})]$ mixing with additional two-meson
interactions.}
\begin{tabular}{c|c|cc}
\hline \hline $(N^{*P},\Delta^{P^{'}})$ & ($N, ~~N^{*}$) & $\Delta$
(MeV) & $\Delta_{\rm expt.}$ (MeV)  \\
\hline $(-,+)$ & N(940), R(1535) & 2330 & 1910 \\
$(+,-)$ & N(940), R(1440) & 2030,2730 & - \\
$(-,-)$ & N(940), R(1535) & 1140 & -  \\
 \hline \hline
\end{tabular}
\label{tab:prediction3a}
\end{center}
\end{table}
%=============================================

%===============================================================
\subsection{Masses due to $[({\bf 6},{\bf 3})\oplus({\bf
3},{\bf 6})]$--$[(\mathbf{1},\mathbf{8})\oplus(\mathbf{
8},\mathbf{1})]$--$[(\mathbf{3},\mathbf{\bar 3})\oplus(\mathbf{ \bar
3},\mathbf{3})]$ mixing} \label{ssect:(6,3)(3,3)masses}
%===============================================================

To improve our analysis, we can add a third field, and altogether we
consider
\begin{eqnarray}
\nonumber B_1 &\in& (\mathbf{1},\mathbf{8}) \oplus (\mathbf{
8},\mathbf{1}) [{\rm mir}] \, ,
\\ (B_2,\Delta) &\in& ({\bf 6},{\bf 3})\oplus({\bf 3},{\bf 6}) \, ,
\\ (B_3,\Lambda) &\in& ({\bf 3},{\bf\bar 3})\oplus({\bf\bar 3},{\bf 3}) \,
.
\end{eqnarray}
As discussed above, the case IV-I is possible to produce the
experimental couplings $g_A^{(0)}$ and $g_A^{(3)}$, although this is
$U_A(1)$ violated. To study this mixing, we need to use the previous
Lagrangian Eq.~(\ref{def:18}) as well as the new ones:
\begin{eqnarray}\label{def:18_2}
\nonumber \mathcal{L}^\prime_{(9)} &=& - g_4 \Big ( \bar B_3 \sigma
B_3 - 2 \bar \Lambda \sigma \Lambda \Big ) + \cdots \, ,
\\ \mathcal{L}_{(B)} &=& - g_5 \bar B_1 \sigma B_3  + \cdots \, ,
\end{eqnarray}
that follow from Eqs. (\ref{def:9 int}) and (\ref{e:8/9b interact}).
Chiral symmetry is spontaneously broken through the ``condensation''
of the sigma field $\sigma \to \sigma_0 = \langle \sigma \rangle_{0}
= f_{\pi}$, which leads to the dynamical generation of baryon
masses:
\begin{eqnarray}
\mathcal{L} &=& - f_\pi (\bar B_1, \bar B_3 , \bar B_2) \left(
\begin{array}{ccc} g_1 & g_5 & g_3
\\ g_5 & g_4 & 0
\\ g_3 & 0 & g_2
\end{array} \right ) \left(
\begin{array}{c} B_1
\\ B_3
\\ B_2
\end{array} \right ) + 2 g_4 f_\pi \bar \Lambda \Lambda + 2 g_2 f_\pi \bar \Delta^i \Delta^i
\end{eqnarray}
Since $g_6$ of $g_6 \bar B_3 B_2$ vanishes, we only need five
conditions to solve this system. Therefore, we may use the three
lowest-lying nucleon states $N(940)$, $N(1440)$ and $N^*(1535)$ as
well as the two mixing angles $\theta ^o= 50.7^o$ and $\phi =
66.1^o$. Finally we find that there are two real possibilities as
shown in Table~\ref{tab:prediction4a}.
%=============================================
\begin{table}[tbh]
\begin{center}
\caption{The values of the $\Delta$ and $\Lambda$ baryon masses
predicted from the isovector axial coupling $g_{A~\rm mix.}^{(1)}
= g_{A~ \rm expt.}^{(1)} = 1.267$ and $g_{A~\rm mix.}^{(0)} = 0.33
\pm 0.08$ and the mass fit to $N(940)$, $N(1440)$ and
$N^*(1535)$.}
\begin{tabular}{c|ccccc|cc}
\hline \hline No. & $g_1$ & $g_2$ & $g_3$ & $g_4$ & $g_5$ &
$\Lambda^P$ (MeV) & $\Delta^P$ (MeV) \\
\hline 1 & $4.6$ & 8.0 & $-1.8$ & $-6.1$ & 9.7 & $1580^+$ & $2070^-$
\\ 2 & $-8.4$ & 4.3 & 7.1 & 10.6 & $-2.4$ & $2750^-$ & $1124^-$
\\ 3 & $-1.3$ & 10.2 & 2.1 & $-2.5$ & 9.8 & $640^+$ & $2660^-$
\\ 4 & $-8.7$ & 8.1 & 7.3 & 7.1 & 2.9 & $1850^-$ & $2110^-$
\\ \hline \hline
\end{tabular}
\label{tab:prediction4a}
\end{center}
\end{table}
%=============================================
Once again, the two odd-parity $\Delta$ options appear as the best
ones. First, even-parity flavor-singlet $\Lambda(1580)$, lies very
close to the (three star PDG, Ref. \cite{Amsler:2008zzb})
$P_{01}(1600)$ resonance. Second, the odd-parity flavor-singlet
$\Lambda$ lies at 1850 MeV, also very close to the (three star
PDG, Ref. \cite{Amsler:2008zzb}) $S_{01}(1800)$ resonance. These
are our best candidates in the $[({\bf 6},{\bf 3})\oplus({\bf
3},{\bf 6})]$--$[(\mathbf{1},\mathbf{8})\oplus(\mathbf{
8},\mathbf{1})]$ mixing scenario, that shows that this option is
open.

%===============================================================
\subsection{Baryon masses and chiral restoration}
\label{ssect:chir restor}
%===============================================================

Note that, starting from the above mass formulas one may study the
behavior of baryon masses in the chiral restoration limit, i.e.,
as $f_{\pi} \to 0$. We do not wish to go into this subject in any
depth here, except to point out several more-or-less immediate
consequences of our results.

First we note that in the two-flavor case one often finds nucleon
parity doublets in the chiral restoration limit $f_{\pi} \to
0$~\cite{Dmitrasinovic:2009vp}. That, however, is generally a
consequence of the assumptions made about the number and kind of
chiral multiplets that are being mixed: If one assumes, as in our
studies above, that more than two multiplets are mixed, then, of
course, there will be no parity doublets, but triplets, or generally
as many states as there are admixed multiplets. Moreover, if there
are more than two degenerate states, such as in our studies above,
then at least two will have the same parity, i.e. the concept of
``parity doublets'' ceases to be meaningful and ``parity
multiplets'' ought to be introduced. Finally, if two different
flavor $SU(3)$ multiplets form one chiral multiplet, such as the
${\bf 8}$ and ${\bf 10}$ in the $[({\bf 6},{\bf 3})\oplus({\bf
3},{\bf 6})]$, then the two flavor $SU(3)$ multiplets may form a
mass-degenerate ``parity doublet" in the chiral restoration limit,
even though most of the states in such doublets do not have the same
flavor quantum numbers.

Various conjectures have been made about the potential relation
between the observed parity doublets high in the baryon spectrum
and chiral symmetry restoration, especially the restoration of the
(otherwise explicitly broken) $U_{A}(1)$ symmetry (see Ref.
\cite{Jaffe:2006jy} and references therein). Our results above
{\it viz.} that there are two basic allowed scenarios that differ
in the $U_A(1)$ (non)symmetry of their interactions, show
immediately that the $U_{A}(1)$ symmetry need not play a role in
the baryon spectra. In this regard we agree with the conclusions
of Ref.~\cite{Jaffe:2005sq,Jaffe:2006jy}, who used only a
two-flavor model, however. Such conclusions were also previously
reached in the two-flavor case in Ref.~\cite{Dmitrasinovic:2009vy}
and in Ref.~\cite{Christos}, only in the more restricted case of
just one $SU(2)$ parity doublet and without mirror fields. The
first, limited, attempts at the three-flavor case were made in
Refs.~\cite{Christos2,Zheng:1992mn}.

%===============================================================
\section{Summary and Outlook}
\label{sect:summary}
%===============================================================

We have used the results of our previous paper~\cite{Chen:2009sf}
to construct the $SU_L(3) \times SU_R(3)$ chiral invariant
interactions based on the phenomenological facts regarding the
baryon axial currents, of the chiral $[({\bf 6},{\bf
3})\oplus({\bf 3},{\bf 6})]$ multiplet mixing with other
non-exotic baryon field multiplets, such as the
$[(\mathbf{3},\mathbf{\bar 3}) \oplus (\mathbf{ \bar
3},\mathbf{3})]$ and $[(\mathbf{ 8},\mathbf{1}) \oplus
(\mathbf{1},\mathbf{8})]$.

The existence of these multiplets is not limited to three-quark
interpolators: they are present in the the $SU(3)_L \times SU(3)_R$
Clebsch-Gordan series for the 5-quark interpolating fields, as well
as the 7-quark ones, etc.. Indeed, these are the only non-exotic
chiral multiplets, as they consist of only non-exotic flavor $SU(3)$
multiplets. The ``ordinary'' (vector) $SU(3)$ multiplet content of a
chiral multiplet is determined by the Clebsch-Gordan series for the
tensor product of the right- and left- $SU(3)$ multiplets: thus
$\mathbf{1} \oplus \mathbf{8} \in (\mathbf{3},\mathbf{\bar 3});~
\mathbf{8} \in (\mathbf{ 8}, \mathbf{1}); ~\mathbf{8} \oplus
\mathbf{10} \in ({\bf 6},{\bf 3})$. Introducing multiple fields with
identical chiral contents would lead to double counting, however.
That is to say that the effects of multi-quark fields are implicitly
accounted for, unless these fields differ from the ones we assumed
in some respect other than the non-Abelian chiral multiplet.
%e.g. such as the axial baryon number, which they may.
Introduction of exotic chiral multiplets, on the other hand, would
lead to exotic flavor $SU(3)$ multiplets in the spectrum, which are
absent experimentally, however. Thus, we may conclude that these
three chiral multiplets, together with their mirror images, are the
{\it only} ones consistent with the present experimental knowledge,
and that no additional chiral mixing is phenomenologically allowed,
without further explanation.

The results of the three-field (``two-angle'') mixing are curious
insofar as all phenomenologically permissible combinations of
interpolating fields lead to the same $F$,$D$ values, that are in
reasonable agreement with experiment. This (unexpected)
equivalence of results is a consequence of the relation
$g_{A}^{(0)}= 3F-D$ between the flavor singlet axial coupling
$g_{A}^{(0)}$ and the (previously unrelated) flavor octet $F$ and
$D$ values. That relation is a benchmark feature of the
three-quark interpolating fields and any (potential) departures
from it may be attributed to interpolating fields with a number of
quarks that is higher than three.

We constructed all $SU_L(3) \times SU_R(3)$ chirally symmetric
baryon-one-meson interactions that mix the three basic baryon
chiral multiplets (and their mirror images). All of these
interactions, with only one exception, obey the $U_{A}(1)$
symmetry as well. We used these interactions to relate the mixing
angles to the masses of physical (``mixed'') baryons. Then we
tried to reproduce the phenomenological mixing angles based on
observed baryon spectra. Once the number of admixed fields exceeds
three there is too much freedom, i.e. too many mixing angles, in
the most general form of such a mixing procedure to be constrained
by only three measured numbers. That assumption can be relaxed,
if/when more detailed studies become necessary if/when new
observables are measured in the future.

For the purpose of simplification we used the two lowest-lying
nucleon states and then ``fit" the phenomenological values of the
mixing angles and thus predicted (at least) one high-lying
resonance, which we then searched for in the PDG tables, Ref.
\cite{Amsler:2008zzb}. This has led us to (at least) two allowed
scenarios. In this way we have made the first tentative
assignments of observed baryon states to chiral multiplets. As
explained above, this procedure does not necessarily lead to
unique results, however. The two basic allowed scenarios differ
primarily in the number of predicted flavor-singlet $\Lambda$
hyperons and in the $U_A(1)$ (non)symmetry of their interactions.
At this moment in time we have no reason to prefer one solution to
another, other than aesthetical ones, such as the $U_A(1)$
symmetry breaking.

Manifestly, the good $U_{A}(1)$ symmetry limit is sufficient to
reproduce the nucleon axial couplings and the low-lying spectrum,
as shown in the first scenario ($[({\bf 6},{\bf 3})\oplus({\bf
3},{\bf 6})]$ -- $[(\mathbf{\bar 3},\mathbf{3})\oplus(\mathbf{
3},\mathbf{\bar 3})]$ -- $[(\mathbf{3},\mathbf{\bar
3})\oplus(\mathbf{ \bar 3},\mathbf{3})]$ mixing), but it is not
necessary, as shown in the second scenario ($[({\bf 6},{\bf 3})
\oplus ({\bf 3},{\bf 6})]$ -- $[(\mathbf{1},\mathbf{8}) \oplus
(\mathbf{ 8},\mathbf{1})]$ -- $[(\mathbf{3},\mathbf{\bar 3})
\oplus (\mathbf{ \bar 3},\mathbf{3})]$ mixing). This result stands
in contrast to the two-flavor
case~\cite{Dmitrasinovic:2009vp,Dmitrasinovic:2009vy}, where all
$SU_L(2) \times SU_R(2)$ symmetric interactions have both a
$U_A(1)$ symmetry-conserving and a $U_A(1)$ symmetry-breaking
version. Thus, the three-flavor chiral symmetry is more
restrictive and consequently more instructive than the two-flavor
one.

As a simple corollary of this result follows one of our
conclusions: the mass degeneracy of opposite-parity baryon
resonances is not necessarily a consequence of the explicit
$U_{A}(1)$ symmetry restoration in agreement with the conclusions
drawn from the two-flavor model calculations,
Ref.~\cite{Jaffe:2005sq,Jaffe:2006jy}. Moreover, the parity
doubling need be neither one of, nor the only consequence of the
spontaneous $SU_L(3) \times SU_R(3)$ symmetry restoration.

This result also shows that the ``$U_A(1)$ anomaly'' in QCD may
still, but need not be the underlying source of the ``spin
problem''~\cite{Bass:2007zzb}, as was once widely
thought~\cite{Zheng:1991pk}. In all likelihood it provides only a
relatively small part of the solution, the largest part coming
from the chiral structure of the nucleon.

The main line of applications of these results lies in the
non-zero density/temperature physics: all previous attempts, see
Refs.~\cite{Papazoglou:1997uw,Beckmann:2001bu} included only the
$[(\mathbf{3},\mathbf{\bar 3})\oplus(\mathbf{ \bar
3},\mathbf{3})]$ baryon chiral multiplet, which naturally led to
axial couplings that differ from the measured ones. Another step,
left for the future, is to include the explicit chiral symmetry
breaking.

\section*{Acknowledgments}
\label{ack}

We wish to thank Profs. Daisuke Jido, Akira Ohnishi and Makoto Oka
for valuable conversations regarding the present work. One of us
(V.D.) wishes to thank the RCNP, Osaka University, under whose
auspices this work was begun, and the Yukawa Institute for
Theoretical Physics, Kyoto, where it was finished, for kind
hospitality and financial support under the YITP Molecule workshop
``Algebraic aspect of chiral symmetry for the study of excited
baryons'' (11/2-20 (2009)) program. The work of one of us (V.D.) was
supported by the Serbian Ministry of Science and Technological
Development under grant number 141025.

\end{document}